\documentclass[times,twocolumn,final]{elsarticle}

\usepackage{medima}
\usepackage{framed,multirow}

\usepackage{amssymb}
\usepackage{latexsym}

\usepackage{amsmath}
\usepackage{array}
\usepackage{booktabs}
\usepackage{makecell}
\usepackage[font=small,labelfont=bf,labelsep=none]{caption}
\usepackage{stfloats}

\usepackage[ruled]{algorithm2e}

\usepackage{flushend}

\usepackage{url}
\usepackage{xcolor}
\usepackage{hyperref}
\definecolor{newcolor}{rgb}{0.21,0.49,0.66}
\definecolor{Rcolor}{rgb}{1,0,0}

\journal{XXXXXXX}

\begin{document}

\verso{Jing Jiao \textit{et~al.}}

\begin{frontmatter}

\title{USFM: A Universal Ultrasound Foundation Model Generalized to Tasks and Organs towards Label Efficient Image Analysis}

\author[1]{Jing \snm{Jiao}}
\author[2]{Jin \snm{Zhou}}
\author[1]{Xiaokang \snm{Li}}
\author[3]{Menghua \snm{Xia}}
\author[1]{Yi \snm{Huang}}
\author[1]{Lihong \snm{Huang}}
\author[1,4]{Na \snm{Wang}}
\author[5]{Xiaofan \snm{Zhang}}
\author[2]{Shichong \snm{Zhou}}
\author[1,6]{Yuanyuan \snm{Wang}}
\author[1,6]{Yi \snm{Guo}\corref{cor1}}
\ead{guoyi@fudan.edu.cn}
\cortext[cor1]{Corresponding authors}

\address[1]{Department of Electronic Engineering, School of Information Science and Technology, Fudan University, Shanghai, China}
\address[2]{Fudan University Shanghai Cancer Center, Shanghai, China}
\address[3]{Department of Radiology and Biomedical Imaging, Yale School of Medicine, New Haven, CT, USA}
\address[4]{SenseTime Research, Shanghai, China}
\address[5]{Shanghai Artificial Intelligence Laboratory, Shanghai, China}
\address[6]{Key Laboratory of Medical Imaging Computing and Computer Assisted Intervention of Shanghai, Shanghai, China}

\received{** December 2023}

\begin{abstract}
Inadequate generality across different organs and tasks constrains the application of ultrasound (US) image analysis methods in smart healthcare. Building a universal US foundation model holds the potential to address these issues. Nevertheless, the development of such foundational models encounters intrinsic challenges in US analysis, i.e., insufficient databases, low quality, and ineffective features. In this paper, we present a universal US foundation model, named USFM, generalized to diverse tasks and organs towards label efficient US image analysis. First, a large-scale \textbf{M}ulti-organ, \textbf{M}ulti-center, and \textbf{M}ulti-device US database was built, comprehensively containing over two million US images. Organ-balanced sampling was employed for unbiased learning. Then, USFM is self-supervised pre-trained on the sufficient US database. To extract the effective features from low-quality US images, we proposed a spatial-frequency dual masked image modeling method. A productive spatial noise addition-recovery approach was designed to learn meaningful US information robustly, while a novel frequency band-stop masking learning approach was also employed to extract complex, implicit grayscale distribution and textural variations. Extensive experiments were conducted on the various tasks of segmentation, classification, and image enhancement from diverse organs and diseases. Comparisons with representative US image analysis models illustrate the universality and effectiveness of USFM. The label efficiency experiments suggest the USFM obtains robust performance with only 20\% annotation, laying the groundwork for the rapid development of US models in clinical practices.
\end{abstract}


\end{frontmatter}


\section{Introduction}
Ultrasound (US) imaging, recognized for its non-invasive, safe, and widely accessible nature, is instrumental in medical diagnostics and therapeutic interventions \citep{liuDeepLearningMedical2019}. The widespread accessibility of US imaging has catalyzed its integration with artificial intelligence, driving substantial progress in the automatic analysis of US images, such as tissue segmentation \citep{xiaMultilevelStructurepreservedGAN2022}, tumor detection \citep{antropovaDeepFeatureFusion2017}, disease diagnosis \citep{kangThyroidNoduleSegmentation2022}, and treatment planning \citep{Fontanarosa_2015}. These successes are attributed to the synergy of data, models and algorithms, namely, fully annotated datasets for specific organs or diseases, specially designed networks and careful training methods. As US imaging expands to new organs and diseases, there is a growing necessity for a US label efficient model generalized to diverse tasks and organs for rapid adaptation and deployment in medical practice. Addressing this demand will significantly broaden the application of US analysis, facilitating the wider implementation of US models in smart healthcare.

Recently, visual foundation models \citep{yuanFlorenceNewFoundation2021a} have emerged as a focal point in the study of natural images, showing potential for creating generalized models. These models are often developed on large-scale, unlabeled datasets using self-supervised pre-training methods, such as masked image modeling (MIM) \citep{heMaskedAutoencodersAre2022} and contrastive learning \citep{chenImprovedBaselinesMomentum2020a}. Benefiting from pre-training on such extensive datasets, these foundation models can derive valuable knowledge from images. The knowledge can be generalized to various downstream tasks and significantly enhance efficiency in model development with outstanding performance, like image classification \citep{xieSimMIMSimpleFramework2022}, segmentation \citep{luDelvingDeeperData2023}, detection \citep{weiContrastiveLearningRivals2022}, and generation \citep{liMAGEMAskedGenerative2023}.

Inspired by the success in the natural visual domain, introducing foundation models to US images holds promise for the rapid development of models across multiple tasks and anatomical structures. Nonetheless, it is hard to share foundation models from natural to US images due to the inherent differences in imaging principles \citep{zhangChallengesPerspectivesFoundation2024}. Hence, there is an urgent need to develop a foundation model tailored for US images, with comprehensive evaluations of its versatility and adaptability in various downstream tasks.

\begin{figure*}[!ht]
	\centering
	\includegraphics[width=\textwidth]{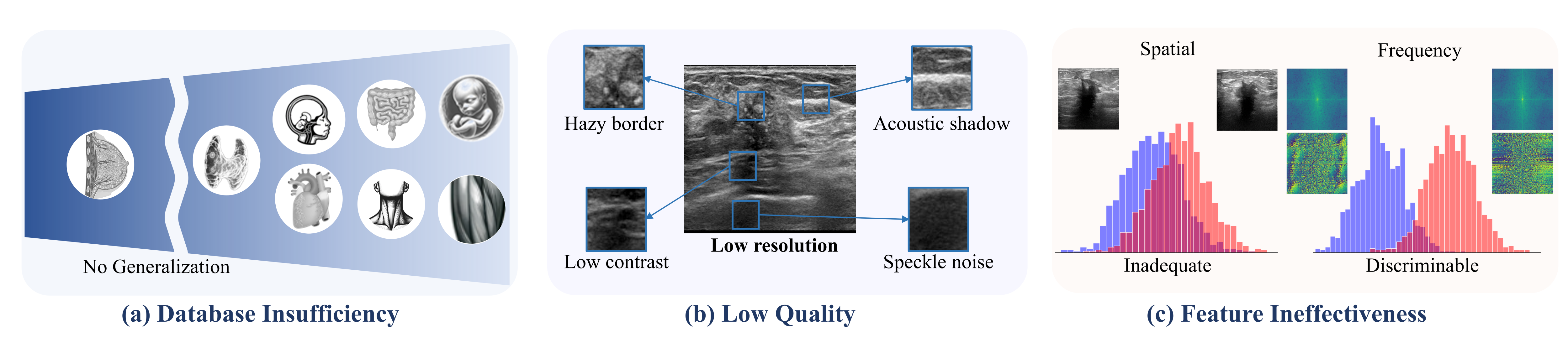}
	\caption{\textmd{Three challenges in US foundation model construction. (a) Database insufficiency. (b) Low quality. (c) Feature ineffectiveness.}}
	\label{fig1}
\end{figure*}

Due to the unique imaging characteristics and data composition of US images, there are three major challenges that need to be addressed in the construction of the US foundation model, as shown in \textcolor{newcolor}{Fig. \ref{fig1}}:
\begin{enumerate}
\item \textbf{Database Insufficiency:} The sufficient and diverse US image database is the basis for a comprehensive US foundation model. Gathering US images from multiple sources that cover a wide range of organs using various imaging devices, is difficult due to complex data acquisition and patient privacy concerns. Additionally, the organs with higher morbidity rates are dominant in the database, leading to organ imbalance.

\item \textbf{Low Quality:} The ability to extract meaningful information robust to noise ensures the applicability of the US foundation model on low-quality US images. US images often have inherent quality issues such as low resolution, contrast, and signal-to-noise ratio (SNR), leading to obscured valuable information. The foundation models built on these noisy and information-sparse US images suffer from difficulty in learning meaningful US information. 

\item \textbf{Feature Ineffectiveness:} Building foundation models capable of extracting effective and general US features is critical for the application of downstream tasks. Extracting clinically relevant and generalizable features from US images is challenging due to the useful US feature is implicit and complex. US features should simultaneously represent the overt spatial features and the underlying practical information concealed within, like the frequency information that depicts the subtle texture changes in the image.
\end{enumerate}

In this study, we have developed a universal US foundation model, namely USFM, with high organ applicability, task adaptability, and label efficiency. To address the above three challenges in US foundation model building, we first established the largest \textbf{M}ulti-organ, \textbf{M}ulti-center, and \textbf{M}ulti-device US database worldwide to date, 3M-US, comprising over two million US images from 12 different human organs. On such a sufficient 3M-US database, we employed an organ-balanced sampling strategy to mitigate organ imbalance. To adequately self-supervised pre-training on low-quality US images, we introduced a novel MIM-based spatial-frequency dual masking method to learn the generalizable and effective US features. The spatial mask enables the USFM to robustly extract useful spatial information from low-quality US images by recovering raw images from the masked (noise-added) inputs. The frequency domain mask is designed to guide the USFM in reconstructing essential frequency information in US images for learning implicit, highly generalizable US representations. The pre-trained USFM can be plug-and-play implemented for various US downstream tasks. Extensive experiments were conducted in our study and have demonstrated its strong generality, superior performance, and excellent label efficiency.

The main contributions of this paper are as follows:
\begin{enumerate}
\item To the best of our knowledge, we are the first to explore and build a US foundation model, namely USFM, which can be used as a plug-and-play module to improve the efficiency and performance of US image analysis across diverse organs and tasks. Facing the insufficient database issue in the development of USFM, we have constructed the largest and most comprehensive 3M-US database to date, including more than two million US images from diverse organs, centers, and devices.

\item To overcome the inherent organ imbalance issue within the 3M-US database, we employed the organ-balanced sampling strategy on the 3M-US database to construct an organ-applicable USFM. Tackling the challenges of low quality and feature ineffectiveness, we introduced a novel MIM-based spatial-frequency dual masking method in the self-supervised pre-training phase of USFM, where spatial masking and frequency masking are proposed to synergistically extract effective US image features, robustly.

\item Massive experiments were conducted to comprehensively validate the robust generalizability and superior performance of the USFM across various tasks and organs. Label efficiency experiments demonstrated that USFM maintains practical performance even with minimal annotated data. Additionally, ablation studies confirmed the effectiveness of our proposed organ-balanced sampling and spatial-frequency dual masking method in addressing the challenges of US foundation model building.

\end{enumerate}

\section{Related work}
\subsection{Visual foundation model}
Inspired by the revolutionary impact of large-scale language models, recent research has extensively focused on large-scale visual foundation models to explore their potential in general vision \citep{chenUnifiedSequenceInterface2022}. These visual foundation models are designed to serve as a universal backbone for various visual tasks, providing a solid foundation for understanding and processing visual data. The universality of these models stems from their pre-training on large-scale, diverse datasets encompassing a broad range of visual content. Visual foundation models can be categorized into two types based on their pre-training approach: task-specific foundation models and task-agnostic foundation models \citep{awaisFoundationalModelsDefining2023}. The former is pre-trained on large annotated datasets to achieve broad applicability for tasks. One of the most representative works is the segment anything model (SAM) \citep{kirillovSegmentAnything2023} developed on a dataset containing one billion labeled segmentation annotations (SA-1B). These task-specific models can be applied through simple prompting or fine-tuning. However, the need for extensive annotated data often limits their development, making them less feasible for tasks with high labeling costs. Considering the abundance of unannotated data, task-agnostic foundational models are established using self-supervised pre-training paradigms to recognize complex visual patterns and learn universal feature representation from larger-scale visual databases. These self-supervised pre-training paradigms are mainly MIM and contrastive learning, where notable works of the former include MAE \citep{heMaskedAutoencodersAre2022}, BEiT \citep{baoBEiTBERTPreTraining2022a}, and the latter includes SimCLR \citep{chenSimpleFrameworkContrastive2020}, MOCO \citep{heMomentumContrastUnsupervised2020}, respectively. These visual foundation models have shown remarkable flexibility and efficiency in various visual tasks, especially in resource-constrained scenarios. The success of these studies paves the way for further exploration and development of advanced visual foundation models tailored to different imaging techniques, holding the promise of advancing the field of visual data analysis and broadening its applications in various contexts.

\subsection{Foundational models in medical imaging}
Despite the remarkable achievements and attention to vision foundation models on natural images, research in the medical domain remains challenging \citep{azadFoundationalModelsMedical2023}. On the one hand, disparities in imaging principles make it troublesome to apply established methods and foundation models for natural images directly to medical images. On the other hand, medical image analysis tasks involve more complex and implicit mapping relationships, requiring foundational models with superior information extraction capabilities. Based on this limitation, many medical imaging foundation models have been established modality-specifically to recognize the significant differences in grayscale distribution and characteristics resulting from the variations in imaging principles. The computed tomography (CT) foundational model, MIS-FM \citep{wangMISFM3DMedical2023}, has been established by pre-training on large-scale 3D volumes and demonstrated its efficacy across multiple target segmentations, including head, neck, thoracic, and abdominal. The foundational model RETFound \citep{zhouFoundationModelGeneralizable2023} has been built by MIM in retinal images and shows high label efficiency in the diagnosis of eye diseases. As for endoscopy videos, a foundational model named Endo-FM \citep{wangFoundationModelEndoscopy2023} has been constructed by contrastive learning and experimented on classification, segmentation, and detection of gastrointestinal diseases. These studies have demonstrated the significant effectiveness of the foundation model in their respective downstream tasks. Given the extensive use of US imaging, developing a universal foundation model for the US will facilitate the advancement of intelligent US analysis in the realm of smart healthcare and broaden its application.

\section{Materials and methods}
\begin{figure*}[!t]
\centering
\includegraphics[width=\textwidth]{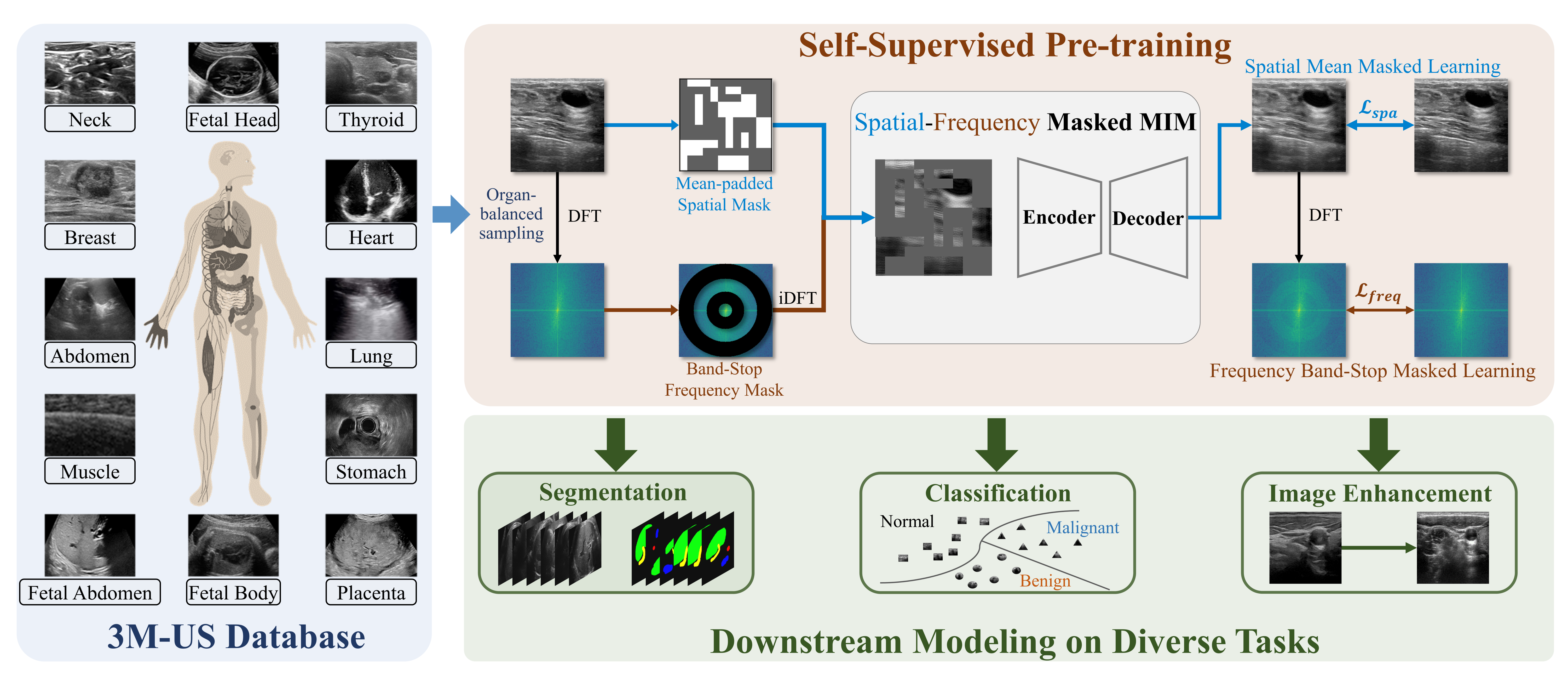}
\caption{\textmd{Overview of the proposed USFM.}}
\label{fig2}
\end{figure*}

\textcolor{newcolor}{Fig. \ref{fig2}} is the overview of our USFM framework. First, a large-scale and comprehensive 3M-US database containing abundant US images from 12 human organs was constructed. Then, images were sampled and fed into the USFM via an organ-balanced sampling strategy to eliminate potential model bias due to data imbalance. The pre-training of USFM, based on the MIM framework, integrates a novel spatial-frequency dual masking scheme. This scheme consists of the spatial mean masking and the frequency band-stop masking. The former simulates common noises in US images, enabling USFM to extract robust and meaningful spatial information. In complementary, the latter facilitates learning of effective implicit features of US images, via recovering masked frequency spectrum. Finally, the pre-trained USFM can be conveniently employed for various downstream tasks involving diverse organs and clinical diseases.

\subsection{The 3M-US database}
\begin{figure}[!ht]
\centering
\includegraphics[width=\linewidth]{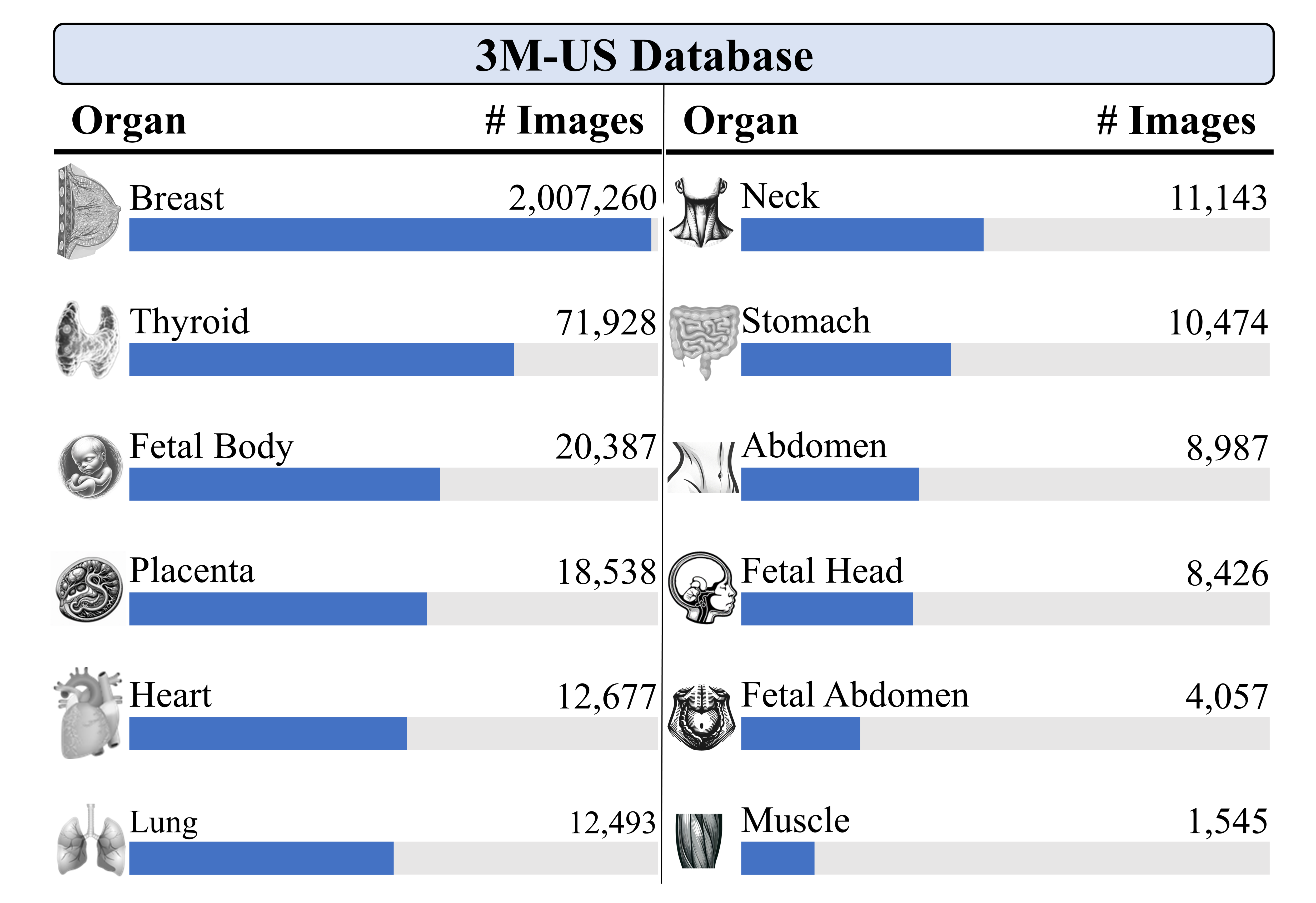}
\caption{\textmd{Summary of the 3M-US database and distribution of involved organs.}}
\label{fig3}
\end{figure}

A comprehensive large-scale dataset is the prerequisite for developing foundation models with general applicability. In our study, we have established the 3M-US, the largest, \textbf{M}ulti-organ, \textbf{M}ulti-center, \textbf{M}ulti-device US database to date. As illustrated in \textcolor{newcolor}{Fig. \ref{fig3}}, the 3M-US database includes 2,187,915 US images of 12 common organs involved in human body screening. The database originates US images from various centers and devices worldwide. By encompassing as many US images as possible, the 3M-US forms the data foundation for achieving organ universality and strong task generalization in our USFM.

All US images included in the 3M-US database underwent the following selection and preprocessing: 1) Images from private databases were used with ethical review approval, and those from public datasets were permitted for public use; 2) Privacy-sensitive information was removed from files and images to ensure data anonymity; 3) All images were converted to grayscale and resized to $(H, W) = (256\times256)$ using bilinear interpolation, then saved in portable network graphic (PNG) format with 8-bit depth (i.e., grayscale images) and no compression to preserve the original intensity values. The private dataset was collected from six medical centers from 2012 to 2022, which included US images of healthy individuals and patients. Most of the images are unlabeled.

The application of US imaging differs markedly depending on the specific organ, resulting in a significant organ imbalance within the 3M-US database, as shown in \textcolor{newcolor}{Fig. \ref{fig3}}. To prevent the model from overlearning the major organs and neglecting the minorities, we employed a weighted organ-balanced resampling strategy. Specifically, a sampling weight $ w_{sample} $ corresponding to different organs was maintained during pre-training on the 3M-US database:
\begin{equation}
    w_{sample}=[\frac{1}{\sqrt{N_{organ1}}},\ \frac{1}{\sqrt{N_{organ2}}},\ ...,\ \frac{1}{\sqrt{N_{organ12}}}],
    \label{eq1}
\end{equation}
where $N_{organ}$ represents the number of images for each organ in the 3M-US database. During each input fetch, the image is sampled from the various organs using weights $w_{sample}$. As a result, images of minority organs will be resampled more frequently to prevent model bias. Moreover, we performed image augmentation on the input data to increase the diversity of the resampled image for model generalization. The augmentation methods include random rotation, scaling, cropping, brightness and contrast adjustment, and Gaussian blur.

\subsection{Spatial-Frequency dual masked image modeling}
In the pre-training phase, we adopt the MIM-based self-supervised pre-training approach to fully utilize the large amount of unannotated data in the 3M-US database, as illustrated in \textcolor{newcolor}{Fig. \ref{fig2}}. The key to functional MIM-based self-supervised pre-training is the optimal design of the masking strategy tailored to the characteristics of the images. For US images, as we have summarized, the characteristics that need to be considered are the low quality and difficulty in effective feature extraction. To address these issues, we propose a spatial-frequency dual masked MIM in the pre-training of USFM, consisting of the spatial mean masked learning and the frequency band-stop masked learning, as illustrated in \textcolor{newcolor}{Fig. \ref{fig4}}.

\subsubsection{Spatial mean masked learning}
Noise in low-quality images, such as acoustic shadows and artifacts, limits robust learning for meaningful US features. To mitigate the impact of noise, we propose spatial mean masked learning in USFM. Unlike existing methods that laboriously eliminate noise, we innovatively continue to add noise to US images in the spatial domain through simple random masking. The USFM is trained in MIM to reconstruct raw images from these masked (noisy-added) images and to gain the ability to recognize noise and robust feature learning.

Specifically, for a US image $U$, we partition it equally into blocks with a given size in the spatial domain. The masking patches are randomly selected based on the masking rate, and their pixel values are replaced. The spatially averaged masking function can be expressed as:
\begin{equation}
    M_{spa}(U)= \begin{cases}
    mean(U),\quad &(x,\ y)\in masked\ patches\\
    0,\quad &otherwise\end{cases}
    \label{eq2}
\end{equation}
where $(x,\ y)$ are the coordinates in the spatial domain, and $mean(U)$ is calculated across the entire image to ensure the continuity of the grayscale distribution. The mask is designed to be mean filling to simulate the noise in US images and avoid the unreasonable disruption of grayscale distribution caused by traditional zero-value (black) filling. Furthermore, USFM is also capable of extracting valuable spatial grayscale and structural information by reconstructing the masked image, such as boundaries, positions, and regional activations.

\subsubsection{Frequency band-stop masked learning}
The extraction of effective US features is fundamental for the high performance of USFM in downstream tasks, while it is often challenging in the spatial domain. US image information in spatial is commonly scarce due to low resolution, low SNR, and sparse information density (limited dynamic range of pixel values). Complementary to the spatial domain, the frequency domain of US images contains richer information, where each component is computed based on global information, reflecting both high-frequency texture variations and low-frequency tissue deformations.

\begin{figure}[!ht]
\centering
\includegraphics[width=\linewidth]{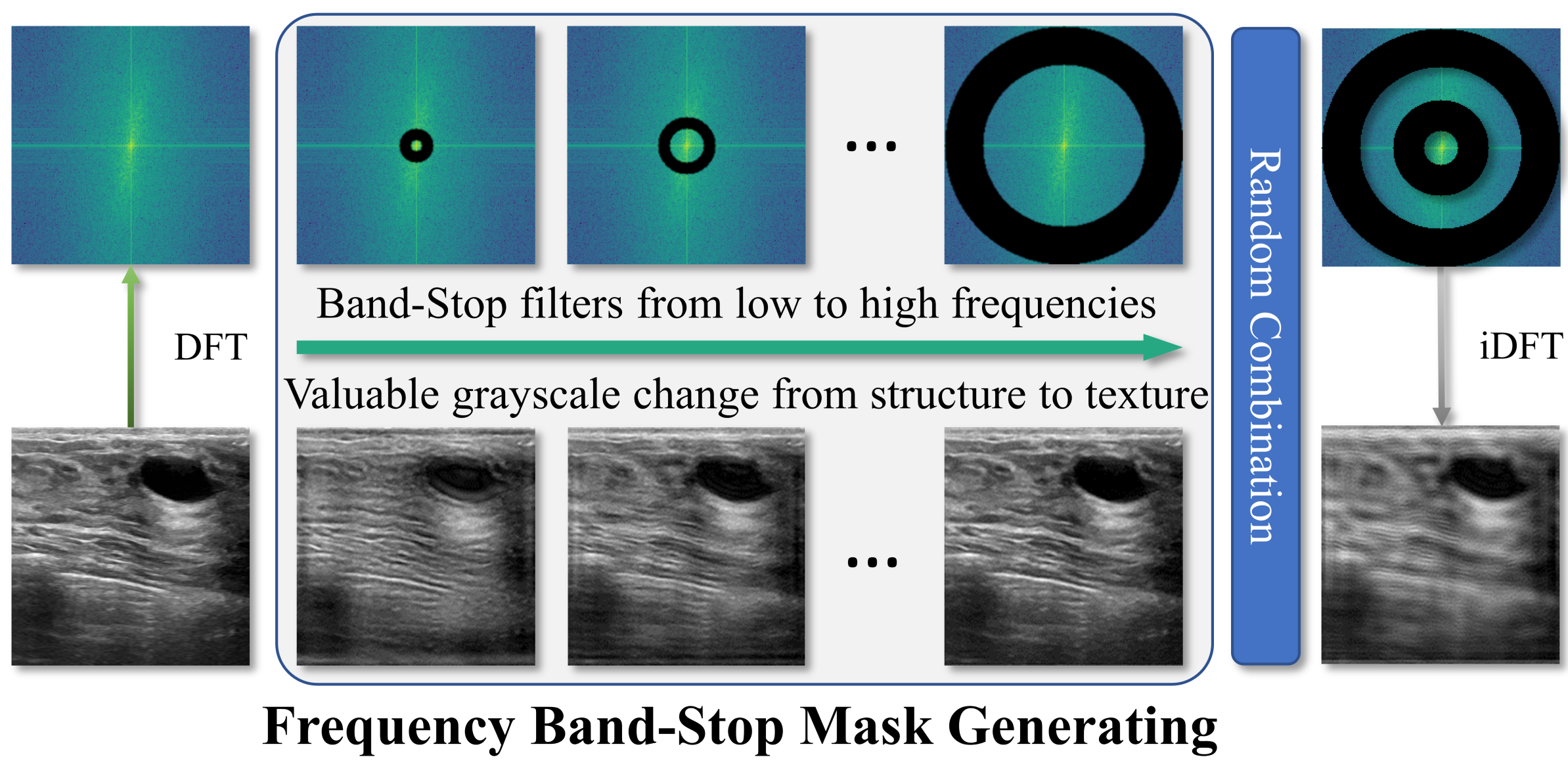}
\caption{\textmd{Illustrate of frequency band-stop mask generating. The frequency domain distribution is obtained from the spatial domain image by DFT, followed by the generation of multiple band-stop filters ranging from low to high frequencies, corresponding to the valuable grayscale change from structure to texture. These filters are randomly sampled and combined to form the final frequency domain mask.}}
\label{fig4}
\end{figure}

As shown in \textcolor{newcolor}{Fig. \ref{fig4}}, the frequency domain distribution of a US image $U$ of size $(H,W)$ can be obtained through 2D Discrete Fourier Transform (DFT):
\begin{equation}
    F(u,v)=\sum_{x=0}^{H-1}\sum_{y=0}^{W-1}U(x,y)\cdot e^{-i2\pi\left(\frac{ux}{H}+\frac{vy}{W}\right)},
    \label{eq3}
\end{equation}
where $(x,y)$ represents the coordinates in the spatial domain, and $(u,v)$ denotes the coordinates in the frequency domain. The $F(u,v)$ represents complex frequency values. According to Euler's formula $e^{i\theta}=\cos{\theta}+i\sin{\theta}$, Equation 3 can also be expressed as:
\begin{small}
\begin{equation}
    F(u,v)=\sum_{x=0}^{H-1}\sum_{y=0}^{W-1}U(x,y)[\cos 2\pi\left(\frac{ux}{H}+\frac{vy}{W}\right) -i \sin 2\pi\left(\frac{ux}{H}+\frac{vy}{W}\right)].
    \label{eq4}
\end{equation}
\end{small}

It is observed that the $F(u,v)$ can be written as $F(u,v)=F_r(u,v)+iF_i(u,v)$ consists of two parts: the real $F_r(u,v)$  and the imaginary $F_i(u,v)$. The amplitude and the phase of the spectrum can be computed from $F(u,v)$ as:
\begin{equation}
    \left|F(u,v)\right|=\sqrt{{F_r(u,v)}^2+{F_i(u,v)}^2},
    \label{eq5}
\end{equation}
\begin{equation}
    \angle F(u,v)=arctan(\frac{F_i(u,v)}{F_r(u,v)}).
    \label{eq6}
\end{equation}

Within the spectrum of a US image, the amplitude and phase indicate the strength and spatial arrangement of various frequency components, respectively. As the shifted amplitude spectrum on the left side of \textcolor{newcolor}{Fig. \ref{fig4}}, the frequency strength gradually goes from low to high from the center outward, and large amplitudes (brighter colors) suggest that these frequency components are vital in US images. The low-frequency components are in the center of the amplitude spectrum, revealing slow-shifted structural information about morphology and deformation, which is essential for US tasks like segmentation and detection. In contrast, the high-frequency components are in the periphery, detailing the rapidly variated textural information and reflecting developmental progressions and pathological alterations, which are vital for US tasks like maturity measurement and tumor staging.

To capture frequency information, we propose a frequency band-stop mask learning method in the MIM framework of USFM. As illustrated in \textcolor{newcolor}{Fig. \ref{fig4}}, the frequency of the US image is randomly masked in the spectrum from low to high frequencies through various band-stop filters. USFM is trained in MIM to extract valuable information across the entire frequency spectrum to recover these masked crucial components. The introduction of frequency band-stop masking enhances the ability of USFM to extract effective US features, which will greatly improve its application in US downstream tasks. For a US image $U$, the frequency band-stop masking function can be represented as:
\begin{equation}
    M_{spa}(U)= \begin{cases}
    0,\quad &f_1 < \sqrt{u^2+v^2} < f_2, ... \\
    F(u,v),\quad &otherwise\end{cases}
    \label{eq7}
\end{equation}
where the $f_1$ and $f_2$ are the lower and upper cutoff frequencies, respectively. As illustrated in \textcolor{newcolor}{Fig. \ref{fig4}}, the final frequency mask $M_{freq}$ is a combination of several band-stop filters to enrich its diversity.

In summary, the spatial-frequency dual masking in the MIM of USFM applies the following operations to the input image $U$, forming a dual-masked input $U_{dm}$:
\begin{equation}
    U_{dm}= \begin{cases}
    iDFT(M_{freq}(DFT(U))),\quad &M_{spa}(U)=0\\
    M_{spa}(U).\quad &otherwise\end{cases}
    \label{eq8}
\end{equation}

That means the spatial and frequency masking are performed on the input US image, respectively. The spatial masked patch in frequency masked image will be replaced with the $M_{spa}(U)$.

\subsubsection{Optimization of self-supervised pre-training}
The self-supervised pre-training of our USFM is achieved by the MIM with spatial-frequency dual masking. During pre-training, the $U_{dm}$ is generated by the dual masking on organ-balanced sampled US image, and then input to the encoder ($E$) and decoder ($D$) structures.
\begin{equation}
    U_{rec}=D(E(U_{dm})),
    \label{eq9}
\end{equation}
and the corresponding reconstructed frequency is:
\begin{equation}
    F_{rec}(u,v)=DFT(U_{rec}).
    \label{eq10}
\end{equation}

The loss of our USFM in the MIM-based self-supervised pre-training phase is
\begin{equation}
    \mathcal{L}_{spa}=\left|U_{rec}-U\right|,
    \label{eq11}
\end{equation}
where $\lambda$ is the scaling factor to adjust the weight of the two reconstruction losses in the training process. The $\mathcal{L}_{spa}$ and $\mathcal{L}_{freq}$ are the reconstruction loss in the spatial and frequency domain.

In the spatial domain, drawing from prior studies on MIM, we introduce an L1 loss for $\mathcal{L}_{spa}$ to supervise USFM restoring the raw image at the pixel level.
\begin{equation}
    \mathcal{L}_{spa}=\left|U_{rec}-U\right|.
    \label{eq12}
\end{equation}

In the frequency domain, the amplitude and phase value ranges across various frequency components exhibit significant variability. Commonly used loss functions, like L1 or L2, typically aim to minimize the overall discrepancy between the raw and reconstructed frequency spectrums. These losses will erroneously bias the USFM to disproportionately prioritize components with larger value ranges while neglecting clinically significant components with smaller value ranges. To address this limitation, we employ a focal frequency loss \citep{jiangFocalFrequencyLoss2021} for $\mathcal{L}_{freq}$.
\begin{equation}
    \mathcal{L}_{freq}=\frac{1}{HW}\sum_{u=0}^{H-1}\sum_{v=0}^{W-1} w(u,v)\left|F_{rec}(u,v)-F(u,v)\right|^2.
    \label{eq13}
\end{equation}

This tailored loss function is specifically designed to improve the retrieval of important information from the frequency domain, by introducing a weight map $w(u,v)$ that assigns weights to each frequency component.
\begin{equation}
    w(u,v)=\left|F_r(u,v)-F_f(u,v)\right|^\alpha,
    \label{eq14}
\end{equation}
where $\alpha$ is the scaling factor for flexibility ($\alpha = 1$ in our experiments). We further normalize the matrix values into the range $[0, 1]$, where the weight 1 corresponds to the currently most lost frequency, and the easy frequencies are down weighted. The specially designed $\mathcal{L}_{freq}$ ensures a more balanced and accurate reconstruction of all relevant frequency components.

\subsection{USFM adaptation in downstream tasks}
\begin{figure}[!ht]
\centering
\includegraphics[width=\linewidth]{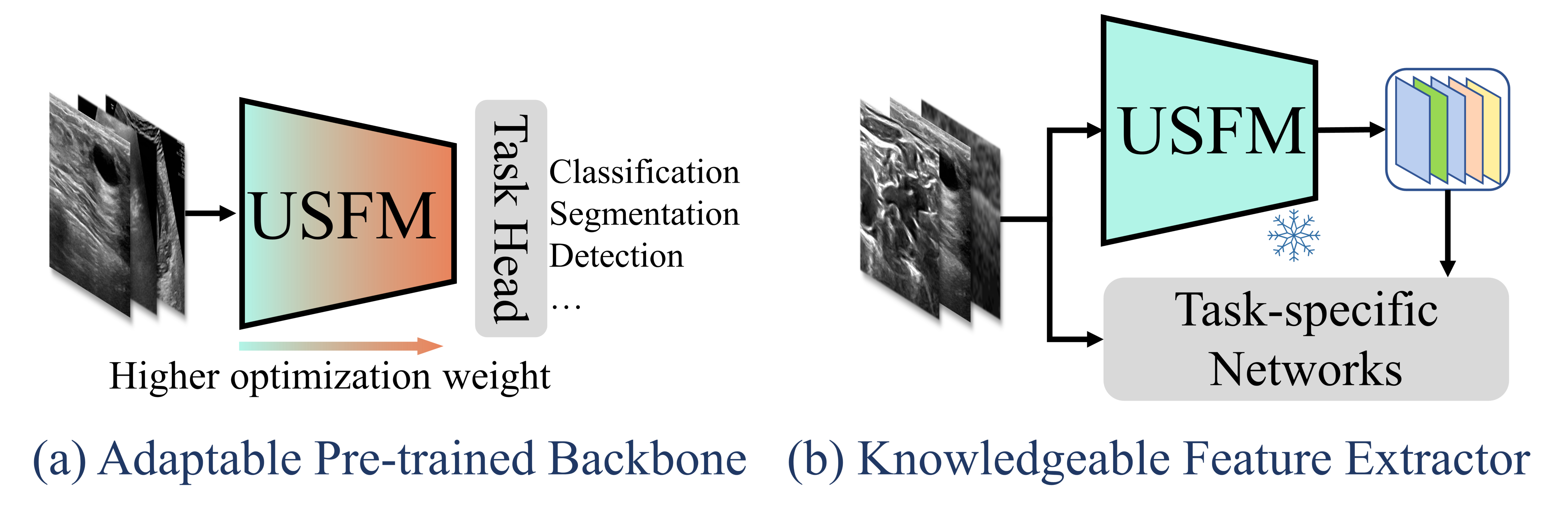}
\caption{\textmd{The diagram of the USFM as a plug-and-play module applied to different tasks: (a) USFM serves as a highly adaptable pre-trained backbone network, which is jointly trained with various task heads. (b) USFM acts as a knowledgeable feature extractor to support task-specific networks.}}
\label{fig5}
\end{figure}

Our USFM can be conveniently used as a plug-and-play module combined with existing methods to achieve better results in downstream tasks. As shown in \textcolor{newcolor}{Fig. \ref{fig5}}, USFM can be integrated with existing methods in two ways.

\textbf{Adaptable pre-trained backbone}: USFM can act as an adaptable pre-trained backbone combined with different task heads to accomplish various downstream tasks. Leveraging its ability to extract universal features learned from the 3M-US database, USFM facilitates rapid fine-tuning with high performance in downstream tasks. In this scenario, USFM is designed for optimization at all layers with incremental weights, whereby shallower layers have smaller optimization weights to ensure stability in extracting low-dimensional spatial features, as low-dimensional spatial information in US images is sparse and changes slowly. Deeper layers have larger optimization weights to accommodate the rapid changes in high-dimensional semantic features.

\textbf{Knowledgeable feature extractor}: USFM can be integrated into them as a knowledgeable feature extractor for the downstream tasks requiring complex or specifically designed network structures. In this case, the weights of USFM are frozen and do not participate in network optimization. The performance of the task network will be enhanced by the stable and effective feature extracted in USFM based on the comprehensiveness of the database.

\section{Experiments}
\subsection{Experimental settings}
To comprehensively validate the effectiveness and applicability of our developed USFM across downstream task datasets of various organs and diseases, we conducted a series of experiments from four aspects:
\subsubsection{Visualization of USFM pre-training}
To verify the effectiveness of organ-balanced sampling and spatial-frequency dual masked MIM in pre-training, we presented the masked image and reconstructed results on the spatial and frequency domain of each organ within the 3M-US database. Moreover, we visualized the distribution in USFM feature space, by randomly selecting 100 US images from 12 different organs in the 3M-US database using UMAP \citep{ghojoghUniformManifoldApproximation2023} dimensionality reduction to demonstrate whether the features are distinguishable across organs without collapsing.
\subsubsection{Comparison on diverse downstream tasks and organs}
Experiments for downstream tasks were conducted on multiple organ datasets, covering three common tasks in US image analysis: segmentation, classification, and image enhancement. The details of the downstream task datasets are provided in \textcolor{newcolor}{Table \ref{tab1}}.

\begin{table*}[!ht]
\centering
\setlength{\abovecaptionskip}{0pt}
\setlength{\belowcaptionskip}{10pt}
\caption{Details of the datasets included in our downstream task.}
\label{tab1}
\begin{tabular*}{\textwidth}{m{0.1\textwidth}m{0.29\textwidth}m{0.13\textwidth}m{0.06\textwidth}m{0.3\textwidth}}
\toprule
Downstream & Datasets & Organs & \#Images & Targets \\
\midrule
\multirow{6}{*}{Segmentation}
& Ultrasound Nerve Segmentation \citep{ultrasound-nerve-segmentation} & Neck & 5735 & Brachial Plexus $vs.$ Background \\ \cmidrule{2-5}
& TN3K \citep{gongThyroidRegionPrior2023} & Thyroid & 3493 & Thyroid nodule $vs.$ Background \\ \cmidrule{2-5}
& BUSI \citep{al-dhabyaniDatasetBreastUltrasound2020} & Breast & 780 & Breast tumor $vs.$ Background \\ \cmidrule{2-5}
& Fetal Abdominal Structures Segmentation \citep{dacorreggioFetalAbdominalStructures2023} & Fetus & 1588 & Artery $vs.$ Liver $vs.$ Stomach $vs.$ Vein $vs.$ Background \\
\midrule
\multirow{5}{*}{Classification} & BUSI \citep{al-dhabyaniDatasetBreastUltrasound2020} & Breast& 780& Benign $vs.$ Malignant $vs.$ Normal\\ \cmidrule{2-5}
& USAnotAI \citep{FtsvdUSAnotAIOrgan}  & Abdomen & 366& Bladder $vs.$ Bowel $vs.$ Gallbladder $vs.$ Kidney $vs.$ Liver $vs.$ Spleen\\ \cmidrule{2-5}
& Fetal Planes \citep{burgos-artizzuEvaluationDeepConvolutional2020a} & Fetus & 12400& Abdomen $vs.$ Brain $vs.$ Femur $vs.$ Thorax $vs.$ Maternal cervix $vs.$ Other \\
\midrule
\begin{tabular}[c]{@{}l@{}}Image\\ Enhancement \end{tabular} & USEnhance2023 \citep{USenhancement2023}& Carotid, Kidney, Liver, Thyroid & 2479 & High quality US images \\ 
     
\bottomrule
\end{tabular*}
\end{table*}

\textbf{Segmentation}: The segmentation experiments were carried out on four datasets of different organs, including the single-object tasks on Neck (brachial plexus), Thyroid (nodule), and Breast (tumor), as well as a multi-object task on Fetus (abdominal structures of artery, liver, stomach, vein, and Background), as shown in \textcolor{newcolor}{Table \ref{tab1}}. USFM was generalized to segmentation tasks as an adaptive pre-trained backbone (\textcolor{newcolor}{Fig. \ref{fig5}}(a)) by adding a segmentation task head, UperNet \citep{xiaoUnifiedPerceptualParsing2018}, as suggested in BEiT \citep{baoBEiTBERTPreTraining2022a}. We compared the segmentation performance of USFM with common US image segmentation methods, including CNN-based Unet \citep{ronnebergerUNetConvolutionalNetworks2015} and ResUnet \citep{zhangRoadExtractionDeep2018}, and vision transformer (ViT) \citep{dosovitskiyImageWorth16x162021} based SegFormer \citep{safaSegFormerSemanticSegmentation2023}. To validate the effectiveness of US image knowledge learned by USFM, UperNet without pre-trained weights and foundation model (FM) based SimMIM \citep{xieSimMIMSimpleFramework2022} trained on natural images were also included in the comparison. Common metrics such as Dice similarity coefficient (DSC), Hausdorff distance at 95th percentile (HD95), intersection over union (IoU), accuracy (ACC), and sensitivity (SEN) were used to report segmentation performance.

\textbf{Classification}: Classification experiments were conducted on the multi-class tasks of Breast (tumors of benign and malignant, and normal), Abdomen (organs of bladder, bowel, gallbladder, kidney, liver, and spleen), and Fetus (planes of abdomen, brain, femur, thorax, maternal cervix, and other), as shown in \textcolor{newcolor}{Table \ref{tab1}}. USFM was generalized to classification tasks by adding a classification task head, i.e., a linear classification layer, following the adaptive pre-trained backbone (\textcolor{newcolor}{Fig. \ref{fig5}}(a)). The commonly used models in medical classification, like ResNet50 \citep{heDeepResidualLearning2016} and DenseNet121 \citep{queDenselyConnectedConvolutional2018}, were included in the comparison. Similar to the segmentation experiments, ViT \citep{dosovitskiyImageWorth16x162021} without pre-trained weights and FM-based SimMIM trained on natural images were also included to demonstrate the role of US image pre-training in USFM. The classification performance was reported using metrics including ACC, recall (Recall), precision (PREC), F1-score (F1), and Matthew’s correlation coefficient (MCC).

\textbf{Image Enhancement}: Enhancement refers to generating high-quality images from low-quality US images using generative models (e.g., CycleGAN \citep{CycleGAN2017}, the most representative). Accurate image enhancement relies on the model learning essential and practical features in US images. In our experiments, USFM was used as a knowledgeable US feature extractor (\textcolor{newcolor}{Fig. \ref{fig5}}(b)) to assist CycleGAN. The image features extracted by USFM were combined with the encoder features of CycleGAN on the channel level to achieve more substantial image encoding capabilities. The original CycleGAN and the CycleGAN with the natural image feature extractor of SimMIM were included in the comparison. The dataset, USEnhance2023, containing low and high-quality images of the Thyroid, Carotid, Liver, and Kidney, was adopted for the experiment. Metrics of structural similarity index measure (SSIM), local normalized cross-correlation (LNCC), and normalized mutual information (NMI) were employed to evaluate the US enhancement performance.

In the downstream tasks of segmentation, classification, and image enhancement, all methods are established on the training set and validated on the validation set. The model demonstrating optimal validation performance is saved and evaluated on the test set. The test results are reported as the mean and standard deviation (mean ± std).

\subsubsection{Validation of label efficiency}
The label efficiency experiment aimed to confirm that constructing US models with USFM can substantially reduce the need for annotation on the downstream tasks, thus providing a fundament for the low-cost development of US models. The label efficiency experiment was conducted on the segmentation and classification tasks, as they usually have high annotation requirements. In the experiment, we randomly sampled 20\%, 40\%, 60\%, and 80\% data from training sets to develop the models on the downstream tasks. The sampling indices on training sets remain the same across different methods for a fair comparison. These models were tested on the entire test set to compare with models developed on the full training set.

\subsubsection{Ablation study}
The ablation experiment was conducted to validate the role of the proposed organ-balanced sampling and spatial-frequency dual masked MIM modeling in establishing USFM. USFM-woS and USFM-woF represent the USFM conducted without organ-balanced sampling and frequency masking during pre-training, respectively. Their performance on downstream tasks was tested and compared with USFM.

\subsection{Implementation details}
During the pre-training phase, US images in the 3M-US database were input at the size of $224\times224$, with the probability of spatial domain masking set at 0.4. For frequency masking, we established seven band-pass filters from low to high frequency and randomly combined two of them as the frequency mask. It is important to note that a $10\times10$ area at the center of the spectrum, containing critical low-frequency information, was always preserved. The balance factors for the two reconstruction losses in spatial and frequency domains were set at 0.4 through grid search.

For the downstream tasks, all images were resized to 256x256 for model input. Basic morphological transformations, random rotations, flips, translations, crops, and scaling were used as data augmentation for the segmentation tasks. As for the classification tasks, only random rotations and flips are used. There is no data augmentation in the US enhancement task. All ViT-based models are implemented by the ViT-b for a balance between performance and efficiency. All models were trained for 200 epochs using the Adam optimizer with a cosine learning rate adjustment strategy. All experiments were conducted on an AMD EPYC 7763 CPU and NVIDIA® GeForce Tesla A100. All models were developed using PyTorch.

\begin{figure*}[!ht]
	\centering
	\includegraphics[width=.75\textwidth]{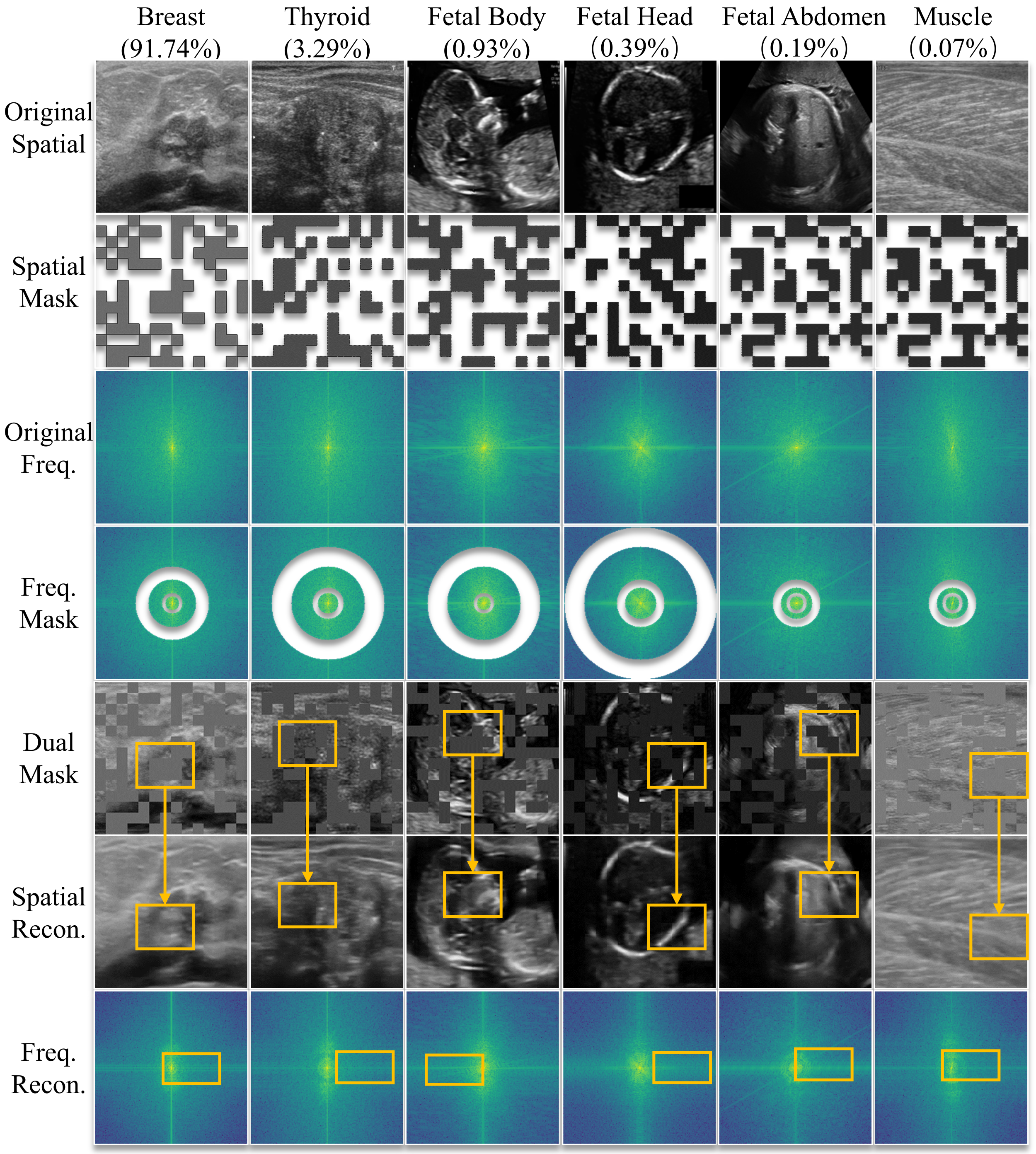}
	\caption{\textmd{Examples of the original image, mask, and reconstruction (Recon.) in the spatial and frequency (Freq.) domain of the organs in the 3M-US database. The yellow box highlights the severely masked structure in the US image and the effective reconstruction of the USFM.}}
	\label{fig6}
\end{figure*}

\begin{figure}[!h]
	\centering
	\includegraphics[width=0.7\linewidth]{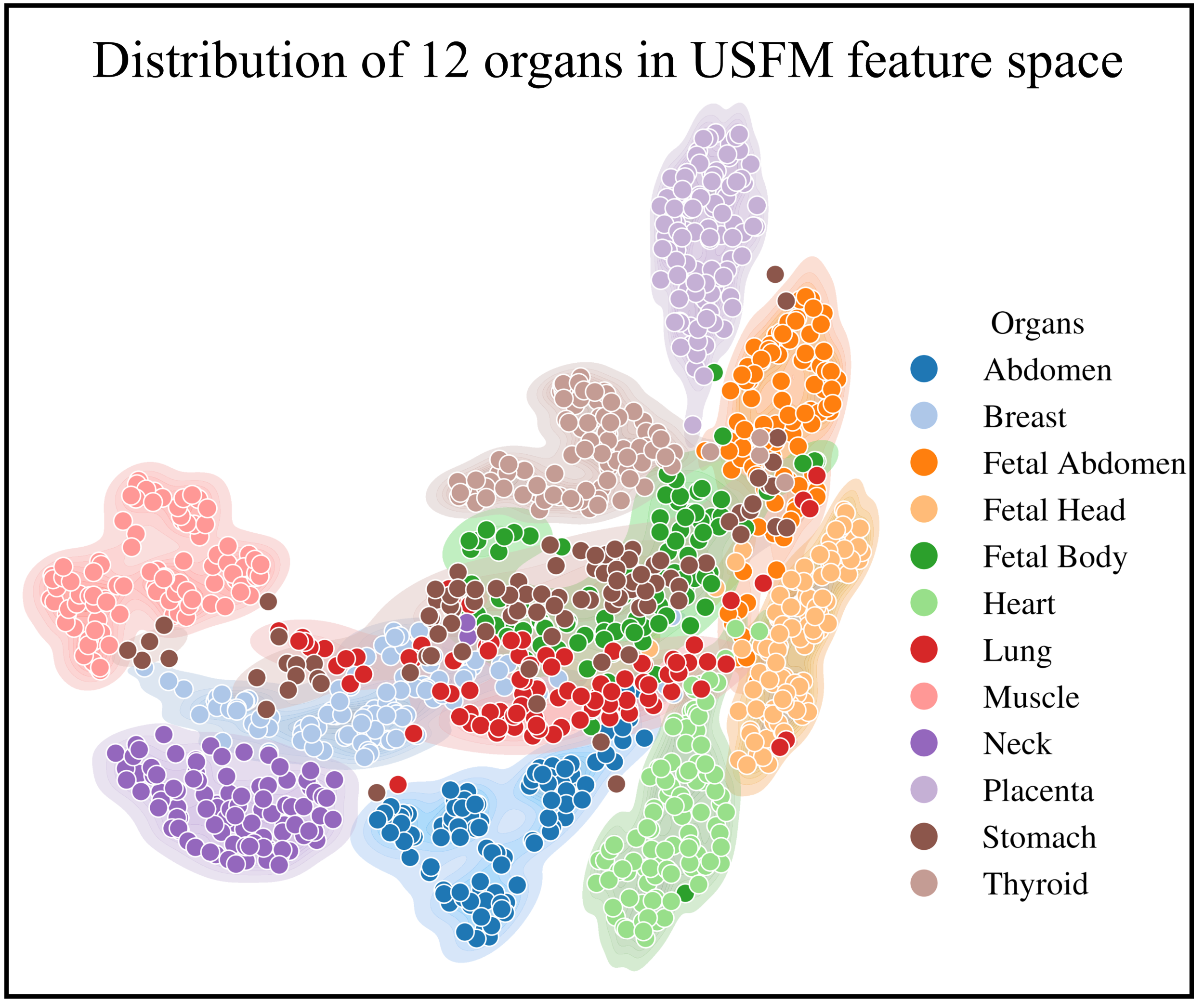}
	\caption{\textmd{Visualization of the UMAP distribution of the 12 organs in the USFM feature space.}}
	\label{fig7}
\end{figure}

\section{Results}
\subsection{Effectiveness of USFM pre-training}
\textcolor{newcolor}{Fig. \ref{fig6}} illustrates the spatial and frequency mask and reconstruction results of our MIM-based USFM in the 3M-US database. The excellent reconstruction results in spatial and frequency domains demonstrate that USFM can effectively learn and reproduce useful features. Thanks to the organ-balanced sampling strategy, USFM maintains a consistent recovery ability for both the majority organs of Breast (91.74\%) and Thyroid (3.29\%), as well as the minority of Fetal Body (0.93\%), Fetal Head (0.39\%), Fetal Abdomen (0.19\%), and Muscle (0.07\%). The ability of USFM to extract organ-universal features lays the foundation for its broad applicability in diverse organ contexts.

The spatial mean mask allowed USFM to counteract the inherent noise in US images by reconstructing the original image from artificially introduced noise. As depicted in the yellow box in \textcolor{newcolor}{Fig. \ref{fig6}}, USFM can successfully recover from noise even in images where critical structures are obscured by noise, such as the masked tumors in Breast, the grayscale distribution in Thyroid and Muscles, and borders in Fetal Body, Head, and Abdomen. This demonstrates the robust feature extraction ability of USFM in low-quality US images.

Frequency masks enable the USFM to extract more effective US features. USFM can recover most of the frequency information in the image even on the aggressive bands-top masking strategy, as shown in \textcolor{newcolor}{Fig. \ref{fig6}} (Freq. Mask $vs.$ Freq Recon.). This indicates that USFM can effectively capture useful frequency information in the image. The spatially reconstructed image benefits from frequency information and exhibits finer grayscale variations (more detailed texture). Extraction of US features from the frequency domain compensates for the limitation of scarce spatial information, which is especially helpful on downstream tasks that require a high-level understanding of US image and texture detail focus.

\textcolor{newcolor}{Fig. \ref{fig7}} visualizes the distribution of the 12 organs from the 3M-US database in the USFM feature space. The samples of each organ are clustered in the feature space of USFM. The clustering suggests that USFM can capture the differences in structural and grayscale information from various organs, which is effective for the US downstream. Furthermore, fetal organs, Fetal Abdomen, Fetal Head, and Fetal Body are close together on the right side of the distribution. Within each organ, the feature representations of different US images are not overlapping, implying that our spatial-frequency dual masked MIM method can prevent USFM from collapsing, i.e., it learns meaningful features rather than noise, which is no definite difference between the images.

\begin{table*}[!ht]
	\centering
	\setlength{\abovecaptionskip}{0pt}
	\setlength{\belowcaptionskip}{10pt}
	\caption{Comparison results on downstream segmentation tasks.}
	\label{tab2}
	\begin{tabular*}{\textwidth}{m{0.12\textwidth}m{0.12\textwidth}m{0.12\textwidth}m{0.09\textwidth}m{0.09\textwidth}m{0.09\textwidth}m{0.09\textwidth}m{0.09\textwidth}}
		\toprule
		Organs                   & Types                      & Models    & DSC (\%) & HD95      & IoU (\%)  & ACC (\%) & SEN (\%)  \\
		\midrule
		\multirow{6}{*}{Neck}    & \multirow{2}{*}{CNN-based} & Unet      & 70.5±21.0 & 35.6±33.7 & 57.8±21.0 & 98.4±0.9 & 72.2±25.9 \\
		&                            & ResUnet   & 78.7±15.9 & 13.4±12.6 & 67.1±17.1 & 98.8±0.8 & 78.9±18.4 \\ \cmidrule{2-8}
		& \multirow{2}{*}{ViT-based} & SegFormer & 79.1±15.5 & 13.3±12.6 & 67.5±16.9 & 98.8±0.8 & 80.1±18.2 \\
		&                            & UperNet   & 76.6±16.7 & 15.3±13.9 & 64.4±17.8 & 98.7±0.8 & 76.9±20.4 \\ \cmidrule{2-8}
		& \multirow{2}{*}{FM-based}  & SimMIM    & 79.4±15.5 & 13.3±12.3 & 67.9±16.7 & 98.8±0.9 & 79.9±17.7 \\
		&                            & USFM      & \textbf{80.6±14.7} & \textbf{12.3±11.6} & \textbf{69.5±16.3} & \textbf{98.9±0.9} & \textbf{80.7±16.7} \\
		\midrule
		\multirow{6}{*}{Thyroid} & \multirow{2}{*}{CNN-based} & Unet      & 78.6±19.9 & 28.1±28.6 & 68.3±21.8 & 95.4±6.0 & 81.9±21.7 \\
		&                            & ResUnet   & 77.9±20.1 & 29.6±32.2 & 67.4±22.4 & 95.2±6.1 & 82.0±21.5 \\ \cmidrule{2-8}
		& \multirow{2}{*}{ViT-based} & SegFormer & 78.9±19.9 & 27.0±27.4 & 68.6±21.8 & 95.4±6.0 & 81.8±21.7 \\
		&                            & UperNet   & 71.1±22.1 & 58.2±37.2 & 59.1±23.7 & 93.6±7.1 & 78.0±22.4 \\ \cmidrule{2-8}
		& \multirow{2}{*}{FM-based}  & SimMIM    & 83.1±19.1 & 21.0±24.2 & 74.5±21.3 & 96.3±5.5 & 84.1±20.0 \\
		&                            & USFM      & \textbf{85.9±14.6} & \textbf{18.4±24.0} & \textbf{77.5±17.9} & \textbf{96.8±5.0} & \textbf{86.4±16.8} \\
		\midrule
		\multirow{6}{*}{Breast}  & \multirow{2}{*}{CNN-based} & Unet      & 76.3±25.7 & 29.9±35.4 & 67.1±26.8 & 95.5±6.1 & 79.1±25.6 \\
		&                            & ResUnet   & 78.5±22.9 & 25.6±31.3 & 69.1±24.1 & 96.3±5.1 & 79.2±24.0 \\ \cmidrule{2-8}
		& \multirow{2}{*}{ViT-based} & SegFormer & 65.3±28.0 & 41.9±38.3 & 54.1±27.6 & 94.6±6.1 & 63.4±29.4 \\
		&                            & UperNet   & 66.0±28.9 & 43.7±39.3 & 55.3±28.2 & 94.8±5.3 & 66.1±30.4 \\ \cmidrule{2-8}
		& \multirow{2}{*}{FM-based}  & SimMIM    & 80.9±24.0 & 22.3±31.3 & 72.9±25.6 & 96.9±4.6 & 80.9±25.3 \\
		&                            & USFM      & \textbf{84.3±17.8} & \textbf{16.7±20.6} & \textbf{76.0±20.3} & \textbf{97.4±3.5} & \textbf{83.9±21.0} \\
		\midrule
		\multirow{6}{*}{Fetus}   & \multirow{2}{*}{CNN-based} & Unet      & 80.3±9.2  & 20.1±5.6  & 68.0±11.6 & 98.2±0.7 & 79.4±12.3 \\
		&                            & ResUnet   & 81.3±9.9  & 19.5±5.6  & 69.5±12.0 & 98.2±0.7 & 82.5±12.7 \\ \cmidrule{2-8}
		& \multirow{2}{*}{ViT-based} & SegFormer & 74.2±14.7 & 22.0±5.9  & 60.6±14.9 & 97.8±1.0 & 69.0±15.6 \\
		&                            & UperNet   & 77.9±11.5 & 20.6±5.6  & 65.0±13.1 & 97.9±0.9 & 78.5±13.7 \\ \cmidrule{2-8}
		& \multirow{2}{*}{FM-based}  & SimMIM    & 83.7±9.1  & 18.4±4.7  & 72.8±10.5 & 98.5±0.6 & 82.0±10.9 \\
		&                            & USFM      & \textbf{85.8±8.5}  & \textbf{17.0±4.4}  & \textbf{75.9±10.1} & \textbf{98.7±0.6} & \textbf{85.5±9.6} \\
		\bottomrule
	\end{tabular*}
\end{table*}

\subsection{Downstream tasks adaption}
\subsubsection{Segmentation on various organs}
The segmentation performance of USFM for four datasets of different organs is reported in \textcolor{newcolor}{Table \ref{tab2}}. The results show that USFM exhibits superior performance across all metrics in US image segmentation tasks. The CNN-based approach is inferior to USFM in the segmentation tasks of all organs due to the simple network structure that cannot adequately fit the complex structural and textural information in the USFM image. Even by increasing the network structure, the ViT-based SegFormer and UPerNet did not show better performance than the CNN-based methods, with the DSC of UperNet inferior to ResUnet on Neck (76.6\% $vs.$ 78.7\%) and Thyroid (71.1\% $vs.$ 77.9\%), and SegFormer inferior to Unet on Breast (65.3\% $vs.$ 76.3\%) and Fetus (74.2\% $vs.$ 80.3\%). These inferiors are due to insufficient data for training the large ViT-based network, like SegFormer and UPerNet.

Compared to these methods, FM-based USFM and SimMIM perform better in all organs. Notably, USFM achieves significant improvements over UperNet (the same network structure but no USFM weight), with DSC increasing from 76.6\% to 80.6\%, 71.1\% to 85.9\%, 77.9\% to 85.8\%, and 66.0\% to 84.3\% on the Neck, Thyroid, Breast, and Fetus, respectively. This demonstrates that USFM has learned critical US knowledge to enhance downstream segmentation tasks by pre-training on a large-scale 3M-US dataset. Due to the fundamental differences between US and natural images, SimMIM, pre-trained on natural images, also shows improved performance in US image segmentation but remains inferior to the USFM. This comparison demonstrates the necessity of the specifically designed USFM for US images in the US segmentation tasks.

Moreover, as seen in \textcolor{newcolor}{Table \ref{tab2}}, except for USFM, Unet, ResNet, SegFormer, and UperNet, show considerable performance variations across different organs. Among these methods, SegFormer performed best on the Neck (79.1\% DSC) and Thyroid (78.9\% DSC), but failed on the Breast (65.3\% DSC). ResUnet only performed best on the Breast (78.5\% DSC) and Fetus (81.3\% DSC), while its performance on Neck and Thyroid was inferior to SegFormer (78.7\%, 77.9\%, $vs.$ 79.1\%, 78.9\% on DSC). As a result, it is difficult to apply any of them as a segmentation method applied generalized to the various organs and tasks. In contrast, our USFM consistently demonstrates superior segmentation performance across all organs and can be used as a universal US image segmentation method.

\begin{figure*}[!ht]
    \centering
    \includegraphics[width=.85\textwidth]{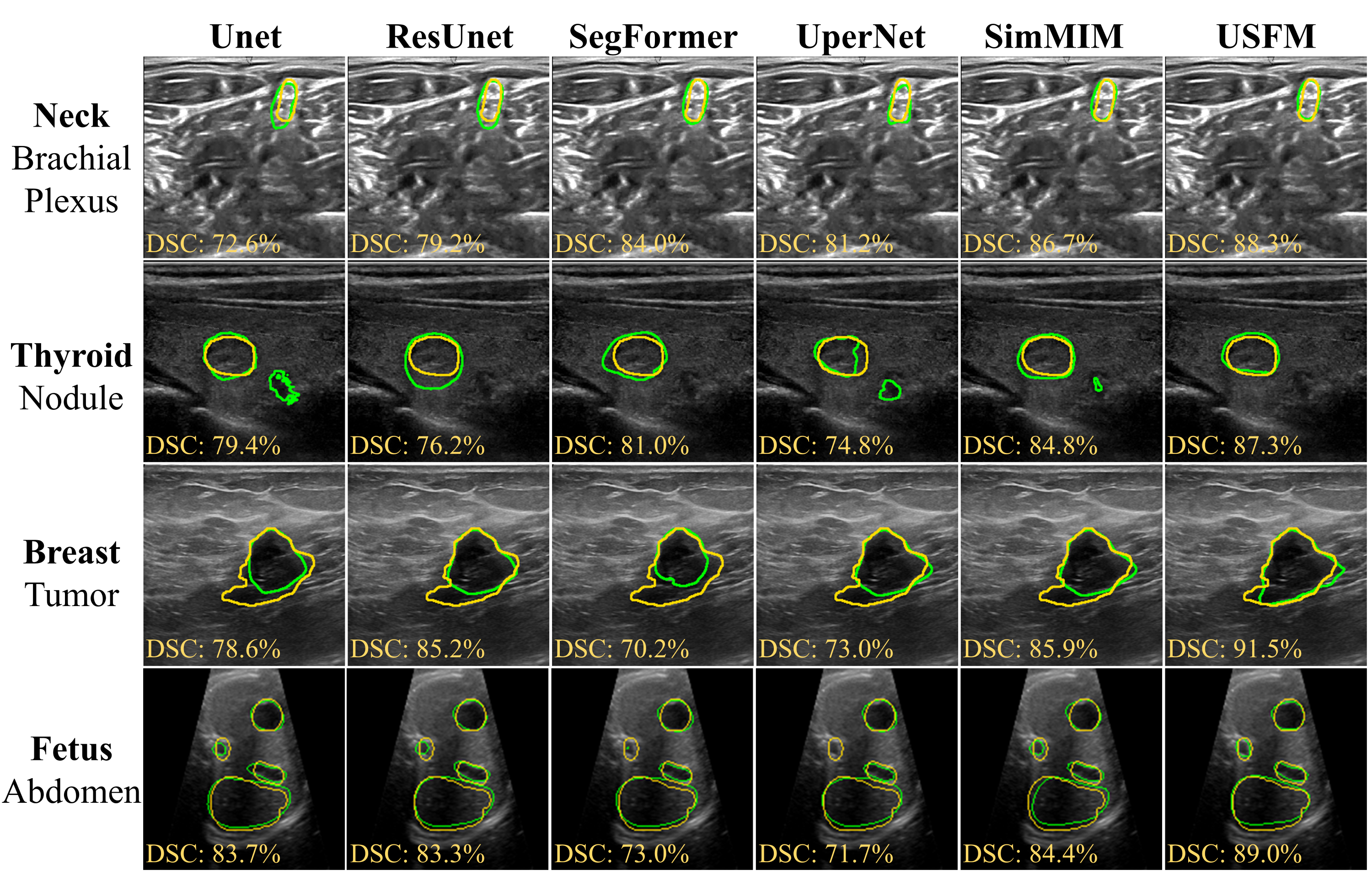}
    \caption{\textmd{The segmentation illustration of the comparison methods. Depicted in yellow is the gold truth, and in green is the predicted result.}}
    \label{fig8}
\end{figure*}

The segmentation results of the comparison methods are illustrated in \textcolor{newcolor}{Fig. \ref{fig8}}. It is evident that USFM outperforms other methods in recognizing the hazy tissue borders and is more robust to noise in the US image. In the segmentation of Neck brachial plexus structure and Breast tumor, Unet, ResUnet, SegFormer, and UperNet under-segment or over-segment due to their limited ability to resolve blurred edges. Moreover, affected by noise in low-quality US images, Unet, UperNet, and SimMIM erroneously segment artifacts in Thyroid nodule segmentation. Compared to methods without pre-training, SimMIM performs better by extracting US structure information shared with natural images. However, its performance is still inferior to USFM due to the lack of adaptation to the distinct characteristics of US image.

\begin{table*}[!h]
	\centering
	\setlength{\abovecaptionskip}{0pt}
	\setlength{\belowcaptionskip}{10pt}
	\caption{ Comparison results on downstream classification tasks.}
	\label{tab3}
	\begin{tabular*}{\textwidth}{m{0.12\textwidth}m{0.12\textwidth}m{0.12\textwidth}m{0.09\textwidth}m{0.09\textwidth}m{0.09\textwidth}m{0.09\textwidth}m{0.09\textwidth}}
		\toprule
		Organs                   & Types                      & Models      & ACC (\%) & Recall (\%) & PREC (\%) & F1 (\%)  & MCC (\%) \\
		\midrule
		\multirow{5}{*}{Breast}  & \multirow{2}{*}{CNN-based} & Resnet50    & 77.3±1.7 & 69.8±2.2    & 78.1±2.5  & 72.8±2.2 & 59.5±3.2 \\
		&                            & DenseNet121 & 85.7±1.6 & 84.6±2.0    & 84.8±1.8  & 84.6±1.8 & 75.3±2.8 \\ \cmidrule{2-8}
		& ViT-based                  & ViT         & 68.8±1.0 & 50.5±1.2    & 49.3±0.9  & 48.2±1.1 & 42.7±2.2 \\ \cmidrule{2-8}
		& \multirow{2}{*}{FM-based}  & SimMIM      & 81.2±1.3 & 79.7±1.4    & 80.2±1.4  & 79.9±1.3 & 67.6±2.2 \\
		&                            & USFM        & \textbf{87.7±1.3} & \textbf{84.4±2.2}    & \textbf{89.0±1.1}  & \textbf{86.1±1.7} & \textbf{78.7±2.4} \\
		\midrule
		\multirow{5}{*}{Abdomen} & \multirow{2}{*}{CNN-based} & Resnet50    & 90.0±1.7 & 90.0±1.7    & 90.7±1.6  & 89.8±1.8 & 88.2±2.0 \\
		&                            & DenseNet121 & 91.7±2.0 & 91.7±2.0    & 94.4±1.4  & 91.6±2.1 & 90.6±2.1 \\ \cmidrule{2-8}
		& ViT-based                  & ViT         & 75.0±2.8 & 75.0±2.8    & 80.2±2.2  & 74.1±3.1 & 71.1±3.2 \\ \cmidrule{2-8}
		& \multirow{2}{*}{FM-based}  & SimMIM      & 90.0±1.9 & 90.0±1.9    & 91.3±1.6  & 89.9±1.9 & 88.3±2.2 \\
		&                            & USFM        & \textbf{93.3±1.8} & \textbf{93.3±1.8}    & \textbf{94.4±1.3}  & \textbf{93.3±1.8} & \textbf{92.2±2.1} \\
		\midrule
		\multirow{5}{*}{Fetus}   & \multirow{2}{*}{CNN-based} & Resnet50    & 83.6±0.2 & 83.3±0.2    & 79.0±0.2  & 80.5±0.2 & 79.7±0.2 \\
		&                            & DenseNet121 & 91.6±0.2 & 92.6±0.2    & 89.1±0.2  & 90.6±0.2 & 89.5±0.2 \\ \cmidrule{2-8}
		& ViT-based                  & ViT         & 79.1±0.2 & 74.9±0.4    & 74.5±0.3  & 74.4±0.4 & 73.5±0.3 \\ \cmidrule{2-8}
		& \multirow{2}{*}{FM-based}  & SimMIM      & 89.7±0.2 & 90.1±0.2    & 86.8±0.2  & 88.1±0.2 & 87.2±0.2 \\
		&                            & USFM        & \textbf{93.6±0.2} & \textbf{94.0±0.2}    & \textbf{91.5±0.2}  & \textbf{92.6±0.2} & \textbf{92.0±0.2} \\
		\bottomrule
	\end{tabular*}
\end{table*}

\subsubsection{Classification of diseases and organs}
The classification performance of USFM in downstream tasks for different organs is reported in \textcolor{newcolor}{Table \ref{tab3}}. Our USFM outperforms other methods across all metrics. It achieved 87.7\%, 93.\%, and 93.6\% ACC on the classification of Breast tumors, Abdomen structures, and Fetus planes, respectively, which is sufficient for clinical practice.

On the classification tasks, the ACC of CNN-based DenseNet121 outperforms ViT on Breast (85.7\% $vs.$ 68.8\%), Abdomen (91.7\% $vs.$ 75.0\%), and Fetus (91.6\% $vs.$ 79.1\%) because its concise network structure can be more adequately trained. The FM-based SimMIM, pre-trained on natural images, was still inferior to DenseNet121 (ACC of 81.2\%, 90.0\%, and 83.6\% $vs.$ 81.2\%, 90.0\%, and 83.6\% on Breast, Abdomen, and Fetus). This result indicates that SimMIM fails to extract useful high-level semantic features in US image classification tasks due to the essential differences between natural and US images. By learning helpful US features from the large-scale US database, our USFM shows a significant performance improvement.

\subsubsection{US image enhancement}
The image enhancement performance on the USEhance2023 dataset is detailed in \textcolor{newcolor}{Table \ref{tab4}}. Compared to the original CycleGAN, using USFM as a knowledgeable auxiliary branch for US features significantly boosts high-quality enhancement performance. USFM achieved superior SSIM compared to CycleGAN: 74.3\% over 72.0\% in the Thyroid, 75.1\% over 73.9\% in Carotid, 76.6\% over 75.8\% in the Liver, and 75.5\% over 75.3\% in the Kidney. This demonstrates the effectiveness of USFM in extracting structural and grayscale information in US images, aiding CycleGAN in achieving better enhancement outcomes. Moreover, USFM achieves superior performance over CycleGAN in LNCC metrics, where are 91.1\%, 89.7\%, 90.3\%, 90.7\% $vs.$ 90.7\%, 89.3\%, 89.2\%, 89.7\% in the Thyroid, Carotid, Liver, and Kidney, respectively. These results demonstrate the ability of the USFM to learn and preserve critical information in US images, ensuring the realism of the generated images (higher LNCC with the target image).

\begin{figure}[!ht]
    \centering
    \includegraphics[width=\linewidth]{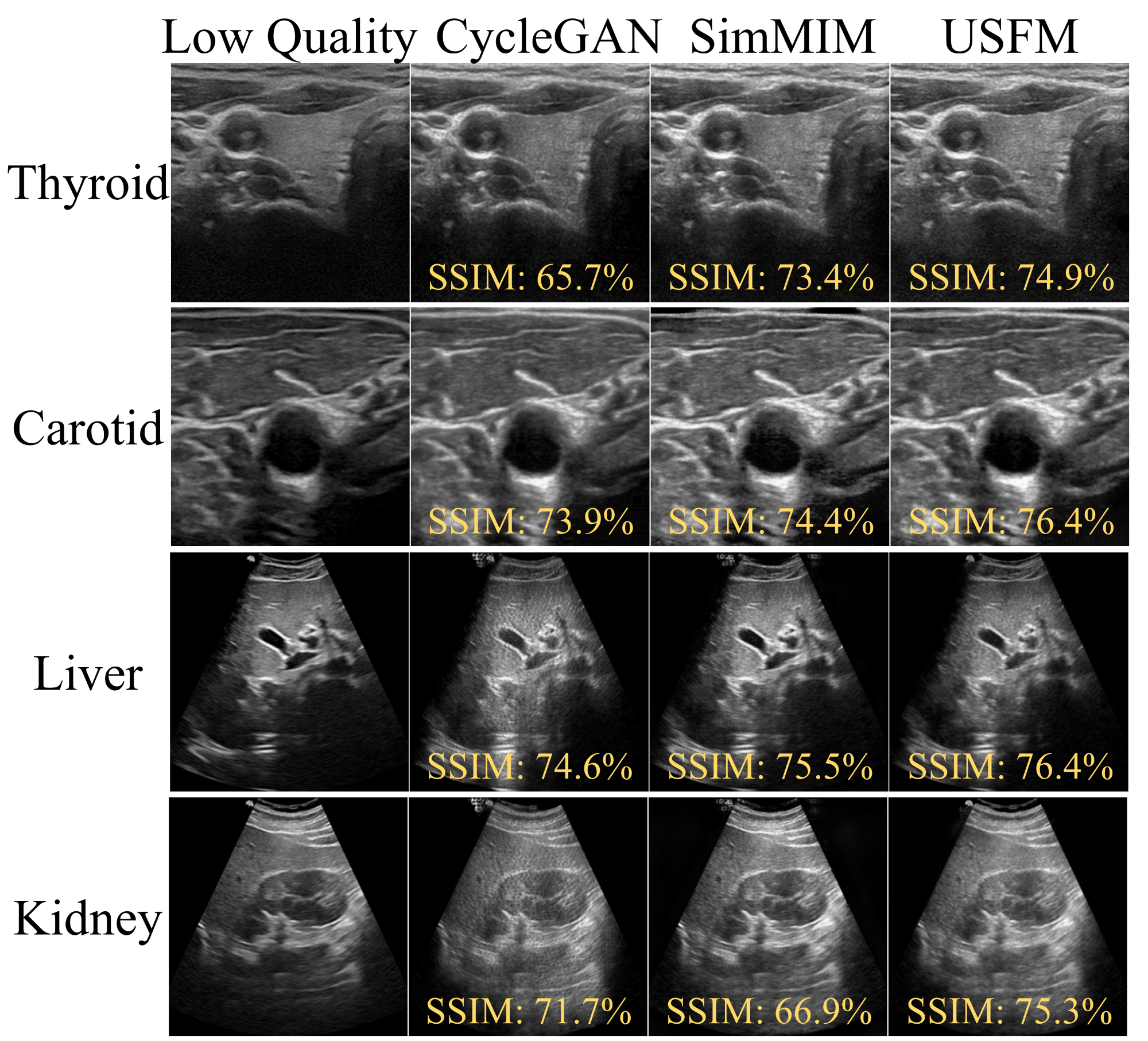}
    \caption{\textmd{An image enhancement example of the comparison methods.}}
    \label{fig9}
\end{figure}

The image enhancement example in \textcolor{newcolor}{Fig. \ref{fig9}} demonstrates that the USFM achieves a higher SSIM than CycleGAN and SimMIM. Without the feature auxiliary branch, CycleGAN displays rough and unrealistic textures in Liver and Kidney. SimMIM alleviates the issue of coarse textures while falling in the image contrast (low contrast between organ and background). In contrast to both CycleGAN and SimMIM, the images enhanced by USFM exhibit finer details and higher contrast, better meeting high-quality US image requirements.
It is worth noting that SimMIM using natural image pre-training as an auxiliary branch did not achieve superior performance in SSIM metrics because of its insufficient ability to extract practical information in US images. On the LNCC metric, SimMIM achieved a weak improvement, benefiting from the understanding of the essential information of the image. Compared to SimMIM, our USFM achieved a substantial lead, further illustrating the superiority of USFM for US images.

\begin{table}[!h]
\centering
\setlength{\abovecaptionskip}{0pt}
\setlength{\belowcaptionskip}{10pt}
\caption{Comparison results on image enhancement tasks.}
\label{tab4}
\begin{tabular*}{\linewidth}{lllll}
\toprule
Organs                   & Models   & SSIM (\%) & LNCC (\%) & NMI (\%) \\
\midrule
\multirow{3}{*}{Thyroid} & CycleGAN & 72.0±3.0 & 90.7±3.4 & 21.6±1.7 \\
                         & SimMIM   & 71.5±3.0 & 90.7±3.7 & 22.3±1.5 \\
                         & USFM     & \textbf{74.3±2.4} & \textbf{91.1±3.3} & \textbf{23.2±2.1} \\
\midrule
\multirow{3}{*}{Carotid} & CycleGAN & 73.9±2.1 & 89.3±4.9 & 21.7±2.0 \\
                         & SimMIM   & 73.5±1.9 & 89.2±5.2 & 21.2±1.5 \\
                         & USFM     & \textbf{75.1±2.1} & \textbf{89.7±4.8} & \textbf{23.0±1.8} \\
\midrule
\multirow{3}{*}{Liver}   & CycleGAN & 75.8±1.8 & 89.2±1.7 & 23.8±1.0 \\
                         & SimMIM   & 75.2±3.1 & 89.5±1.7 & 23.8±1.3 \\
                         & USFM     & \textbf{76.6±2.6} & \textbf{90.3±1.6} &\textbf{ 25.2±1.4} \\
\midrule
\multirow{3}{*}{Kidney}  & CycleGAN & 75.3±1.3 & 89.7±2.3 & 24.5±1.3 \\
                         & SimMIM   & 75.0±2.2 & 89.7±1.9 & 24.2±1.3 \\
                         & USFM     & \textbf{75.5±2.1} & \textbf{90.7±2.0} & \textbf{25.2±1.2} \\
\bottomrule
\end{tabular*}
\end{table}

\begin{figure*}[!ht]
	\centering
	\includegraphics[width=.95\textwidth]{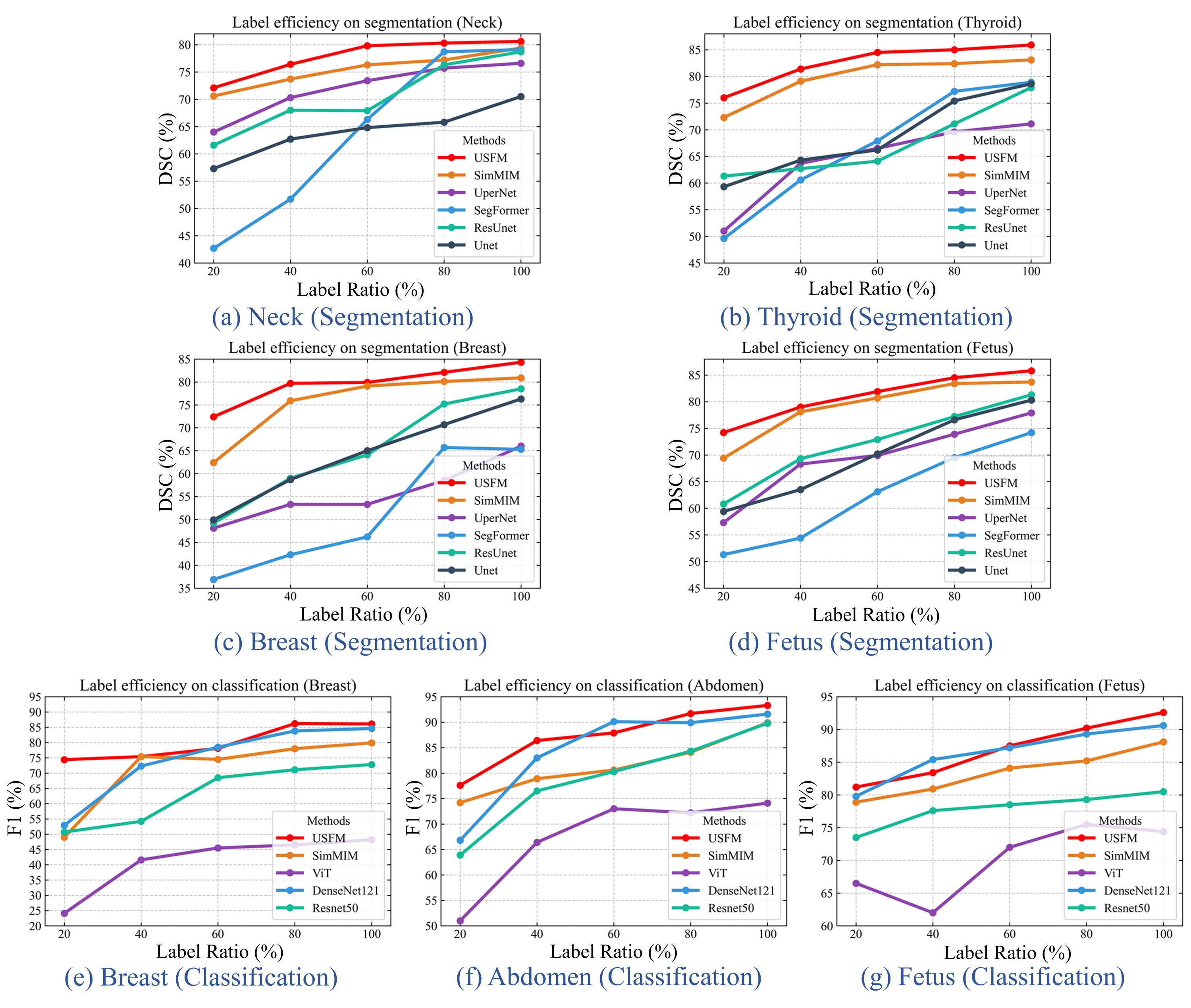}
	\caption{\textmd{Label efficiency experiments of the downstream segmentation and classification tasks. The mean DSC (in segmentation) and mean F1 (in classification) of the model trained at different label ratios are reported.}}
	\label{fig10}
\end{figure*}

\subsection{Label efficiency}
The label efficiency results for the downstream segmentation and classification tasks are presented in \textcolor{newcolor}{Fig. \ref{fig10}}. When developing downstream US on the training set of different label ratios, USFM consistently achieved the highest label efficiency compared to other methods. Even with only a 20\% label ratio available, USFM attain sufficient performance, with DSC of 72.1\%, 76.0\%, 72.4\%, and 74.2\% for the segmentation of brachial plexus in Neck, nodule Thyroid, tumor in Breast, and abdominal structures in Fetus, as well as F1 of 74.4\%, 77.6\%, and 81.2\% for the classification of tumors in Breast, structures in Abdomen and Plane in Fetus. This result indicates that USFM significantly reduces the need for annotations, facilitating the rapid development of US models.

With the label ratio increasing from 20\% to 40\%, USFM achieved substantial improvements in almost all organs. For the segmentation tasks on Neck, Thyroid, Breast, and Fetus, the DSCs have increased from 72.1\% to 76.4\%, 76.0\% to 81.4\%, 72.4\% to 79.7\%, and 74.2\% to 79.0\%, respectively. The F1 of the classification tasks on Abdomen and Fetus have increased from 77.6\% to 86.4\% and 81.2\% to 83.4\%. When the label ratio increased to 60\%, USFM reached near 100\% label ratio performance in the segmentation of Neck, Thyroid, and Breast (\textcolor{newcolor}{Fig. \ref{fig10}}(a, b, c)).

In the segmentation tasks, the without pre-trained Unet, ResUnet, SegFormer, and UperNet, showed strong dependence on the amount of annotated data. At a 20\% label ratio, these four methods failed to achieve acceptable segmentation performance, with DSC below 65\% in Neck, Thyroid, Breast, and Fetus. Except for USFM, all other methods in classification tasks failed to achieve notable performance at a low label ratio. Particularly at the 20\% label ratio, F1 was below 55\% in Breast, 75\% in Abdomen, and 80\% in Fetus. This limitation restricts their application their application in medical practice.

Compared to USFM, SimMIM pre-trained on natural images remained limited in performance due to its weaker feature extraction capability for US images across all label ratios in segmentation tasks. As for the classification tasks, SimMIM performed inferior to the models without pre-training (DenseNet121) at a low label ratio across all organs. This result suggests that SimMIM, pre-trained on natural images, cannot extract high-level US features crucial in US image classification tasks.

\subsection{Ablation analysis}
\begin{figure*}[!ht]
\centering
\includegraphics[width=.9\textwidth]{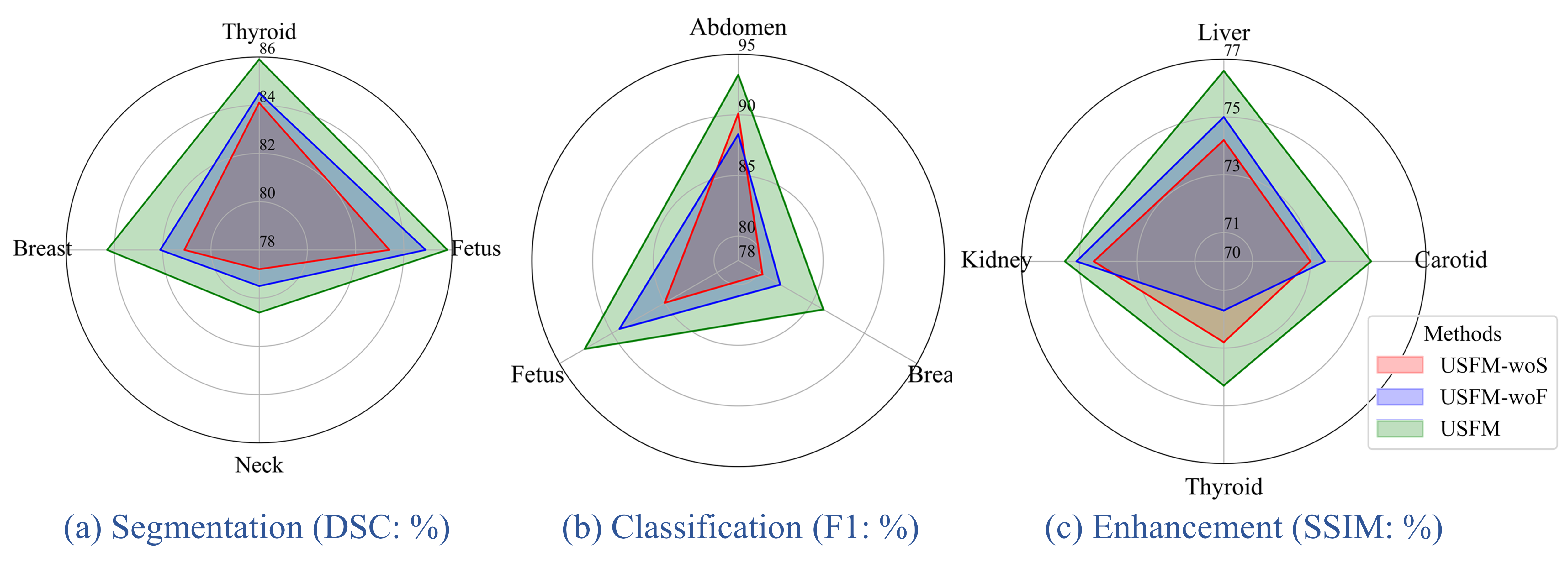}
\caption{\textmd{Ablation study on the segmentation, classification, and detection tasks in all organs. The mean DSC on the segmentation task, mean F1 on the classification tasks, and mean SSIM on the enhancement task are reported to compare USFM-woS, USFM-woF, and USFM.}}
\label{fig11}
\end{figure*}

The performance comparison of USFM, USFM-woS, and USFM-woF in segmentation, classification, and image enhancement tasks is outlined in \textcolor{newcolor}{Fig. \ref{fig11}}. Results show a considerable decrease in performance when our proposed frequency domain masking or organ-balanced sampling is eliminated from USFM.
Due to the removal of organ-balanced sampling, USFM-woS biased the majority organ (Breast, 91.74\%), overfitting meaningless Breast-specific structural details and ignoring the minority organs in the 3M-US database, such as the Thyroid (3.29\%), Fetus (0.93\%), Neck (0.51\%) and Abdomen (0.43\%). The lack of organ applicability in USFM-woS resulted in degraded performance than USFM on all organs, especially in the Breast (-3.2\%) and Thyroid (-1.4\%) on segmentation, Fetus (-6.4\%) and Abdomen (-6.4\%) on classification, and the Carotid (-2.1\%) and Liver (-2.4\%) on image enhancement. Even biased toward Breast, USFM-woS does not perform better than USFM on downstream tasks of Breast, with -3.2\% and -5.2\% degradation on the tumor segmentation and classification. This result indicates that the overfitted Breast features in USFM-woS are not generalizable to the downstream task.
USFM-woF, which removed the frequency band-stop masked learning, focused solely on spatial domain information. Caused by the neglect of implicit frequency information, USFM-woF is insufficient to extract effective US features, thus leading to inferior performance compared to USFM. The inferior was more significant in the classification and enhancement tasks \textcolor{newcolor}{Fig. \ref{fig11}} (b, c), which require a high understanding of the US image. The performance is degraded by -3.9\%, -5\%, and -3.1\% in the classification of Breast tumors, Abdominal structures, and Fetal planes, respectively, and by -1.6\%, -0.4\%, -2.6\%, and -1.6\% in the enhancement of the Liver, kidneys, Thyroid, and Carotid, respectively.

In contrast to USFM-woS and USFM-woF, USFM, by extracting organ-unbiased and effective US features, achieved the best performance across all organs in all tasks. The experimental results further validate the effectiveness of our proposed organ-balanced sampling and spatial-frequency dual masking method in MIM.

\section{Discussion}
The safety imaging principles and the versatile devices make US imaging more widely accessible for disease screening and health management than other medical imaging modalities. In hospital and community healthcare settings, numerous US images are produced in various organs and diseases and require automated analysis. Existing US analysis methods struggle to meet these needs in complex medical scenarios due to insufficient organ applicability, task adaptation, and label efficiency.

USFM, the US foundational model we have built in this study, offers a promising solution to address the shortcomings of current US models and meet the demands in clinical practice. In massive experiments, USFM has demonstrated strong cross-organ applicability, various task adaptability, and label efficiency. As shown in \textcolor{newcolor}{Tables \ref{tab2}}, \textcolor{newcolor}{\ref{tab3}}, and \textcolor{newcolor}{\ref{tab4}}, USFM exhibits robust performance, significantly surpassing representative methods across various organs (Neck, Thyroid, Breast, Carotid, Liver, Kidney, Abdomen, and Fetus) and common US tasks (segmentation, classification, and image enhancement). Moreover, USFM shows exceptional label efficiency in segmentation and classification tasks, as illustrated in \textcolor{newcolor}{Figs. \ref{fig9}} and \textcolor{newcolor}{\ref{fig10}}. Notably, even with only 20\% annotated data, the performance of USFM leads ahead of existing methods trained on complete datasets (superior to UperNet on Neck and Thyroid segmentation, superior to Segformer on Thyroid and Breast segmentation). These excellent properties of USFM stem from our innovative and practical approach to the foundational model construction, addressing three long-standing challenges in the US image: insufficient databases, low quality, and feature ineffectiveness.

Current US models are generally developed on an insufficient database, consisting of images of a particular organ from a single center using the same device. The lack of diversity in training data makes these models unsuitable for complex real-world scenarios, where the US image datasets exhibit significant variability among organs, operator inconsistencies, and equipment deviations. To address this problem, we have established the largest Multi-organ, Multi-center, Multi-device US database, 3M-US, containing over two million US images to cover as many organs, diseases, and centers as possible. Through full training on the large-scale database, USFM is universally applicable to many clinical scenarios. As demonstrated in \textcolor{newcolor}{Table \ref{tab2}}, compared to UperNet with the same network structure but without 3M-US pre-training, USFM shows superior universality across all downstream tasks for all organs. Although SimMIM, pre-trained on natural images, can capture some structural information shared with US images, it fails to extract high-level semantic US features. Therefore, SimMIM consistently underperforms USFM, particularly in classification and enhancement tasks that require a deep understanding of US images. These comparisons validate the comprehensive variety of our 3M-US database, serving as a data foundation for establishing a universal US foundational model.

The superior performance of USFM further demonstrates the relationship between the universality and effectiveness of foundational models. To learn organ-universal features, we employed an organ-balanced sampling strategy in USFM. By balanced learning from different organs in the 3M-US dataset, USFM has adequate feature extraction and reconstruction capabilities for all 12 organs, as demonstrated in \textcolor{newcolor}{Fig. \ref{fig6}}. The effectiveness of these organ-universal features is evidenced in ablation experiments, as shown in \textcolor{newcolor}{Fig. \ref{fig11}}. USFM-woS, removing organ-balanced sampling from USFM, leads to overemphasizing the dominant organs (Breast) during pre-training. The Breast-biased USFM-woS, collapsed to extract Breast features in US images like edges and calcifications, fails in the other organs. This failure significantly degrades performance across all tasks and organs compared to USFM, particularly in classification and enhancement tasks requiring more useful high-level US features. Even on breast tumor segmentation and classification, the organ-universal USFM is more effective than the Breast-biased USFM-woS. The pursuit of universality ultimately leads to enhanced effectiveness, highlighting the significance of developing universal foundational models.

The innovation of USFM also lies in the specially designed spatial-frequency dual masked MIM foundational modeling method, which enables robust learning of effective features on low-quality images. Low quality is an inherent problem in US images, typically manifested as large amounts of noise in low-resolution images, such as acoustic shadows and artifacts. Existing studies primarily focus on image denoising and enhancement. However, the application of these methods on large-scale databases is limited by high computational requirements and the challenges of cross-organ and cross-center adaptation. In contrast to the complex denoising, our USFM proposed a parsimonious and productive spatial mean masking learning method based on MIM. We trained USFM to recover the original images from these artificially masked (noise-added) images. The strong recovery capability, as shown in \textcolor{newcolor}{Fig. \ref{fig6}}, indicates that the features extracted by USFM from the masked (noise-added) images retain the same SNR as the original images, while the most noise is removable masks. As a result, the SNR of the USFM extracted features will be significantly improved with the original unmasked US image input. The robust extraction capabilities of USFM on low-quality noisy US images are demonstrated in \textcolor{newcolor}{Table \ref{tab2}} and \textcolor{newcolor}{Fig. \ref{fig8}}, with superior and stable performance compared to other methods (higher mean and lower std on all metrics). In addition, robustness to noise also allows USFM to learn efficiently from limited data, as the best label efficiency is shown in \textcolor{newcolor}{Fig. \ref{fig10}}. Our proposed MIM-based spatial mask learning approach offers a new direction for future research on low-quality US images.

The inefficiencies of US features are attributed to the scarcity of spatial information and the difficulty of useful knowledge extraction. In the spatial domain, only limited local information is available for US model extraction due to low resolution, low SNR, and low contrast. Concurrently, tasks such as the classification of benign and malignant Breast tumors require a high-level understanding of rapidly changing texture details in US images, which are challenging to capture in the spatial domain. Our USFM employs a frequency band-stop masking learning method within MIM to represent this useful information directly. The frequency domain reconstruction visualization in \textcolor{newcolor}{Fig. \ref{fig7}} shows that USFM can extract useful frequency information from the residual components. The ability to capture frequency domain information allows it to significantly outperform the spatial-only learning USFM-woF across all tasks and organs, as proven in the ablation experiment in \textcolor{newcolor}{Fig. \ref{fig11}}. Especially in classification and enhancement tasks, the neglect of frequency information in USFM-woF cannot extract higher-level semantic features, leading to a substantial decline in performance. By combining spatial mean masking learning and frequency band-stop masking learning, our dual-domain masked MIM approach enables USFM to achieve superior performance and high label efficiency in various tasks and organs, as demonstrated in \textcolor{newcolor}{Tables \ref{tab3}}, \textcolor{newcolor}{\ref{tab4}}, and \textcolor{newcolor}{Fig. \ref{fig10}}.

While our USFM has achieved excellent performance, there is room for improvement. Firstly, our study primarily focuses on 2D US images. Although the analysis of 3D US images can be converted to 2D images, directly establishing a foundational model for 3D US would better utilize the relationships between images. Secondly, limited by computational resources, the potential performance improvements of larger foundational model architectures (ViT-L, ViT-H, Swin Transformer \citep{liuSwinTransformerHierarchical2021a}) remain to be explored. In addition to segmentation, classification, and image enhancement tasks conducted in experiments, USFM also has the potential to be applied to a broader range of US image analysis tasks, such as detection, denoising, and generation. In the future, we will explore practical ways to establish universal foundational models for the US and expand their application scope to achieve widespread use in clinical practice.

\section{Conclusion}
This paper developed a US image foundation model named USFM, characterized by high organ versatility, task adaptability, and label efficiency. The USFM can be a readily plug-and-play module for rapidly expanding automatic analysis models of US. To fulfill the universal applicability of USFM, we established the largest multi-organ US image database to date, named 3M-US, and developed the USFM. Comprehensive experimental results demonstrate that USFM exhibits exceptional performance in various common US tasks across different organs, including segmentation, classification, and enhancement. The outstanding performance of USFM in label efficiency experiments confirms its potential as a foundational model to accelerate the development of US models and promote the expansion of automated US analysis applications.

\section*{Declaration of competing interest}
The authors declare that they have no known competing financial interests or personal relationships that could have appeared to influence the work reported in this paper.

\section*{Acknowledgements}
This work was supported by the National Natural Science Foundation of China (Grant 62371139 and 82227803), the Science and Technology Commission of Shanghai Municipality (Grant 22DZ1100100 and 22ZR1404800).

\bibliographystyle{model2-names.bst}\biboptions{authoryear}
\bibliography{refs}

\end{document}